# Target detection in synthetic aperture radar imagery: a state-of-the-art survey

Khalid El-Darymli
Peter McGuire
Desmond Power
Cecilia Moloney

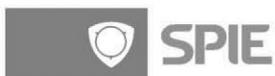

# Target detection in synthetic aperture radar imagery: a state-of-the-art survey

**Khalid El-Darymli,[a,b] Peter McGuire,[a] Desmond Power,[a] and Cecilia Moloney[b]**

[a]C-CORE, Captain Robert A. Bartlett Building, Morrissey Road, St. John's, Newfoundland, A1C 3X5, Canada
Khalid.El-Darymli@mun.ca
[b]Memorial University, St. John's, Newfoundland, A1B 3X5, Canada

**Abstract.** Target detection is the front-end stage in any automatic target recognition system for synthetic aperture radar (SAR) imagery (SAR-ATR). The efficacy of the detector directly impacts the succeeding stages in the SAR-ATR processing chain. There are numerous methods reported in the literature for implementing the detector. We offer an umbrella under which the various research activities in the field are broadly probed and taxonomized. First, a taxonomy for the various detection methods is proposed. Second, the underlying assumptions for different implementation strategies are overviewed. Third, a tabular comparison between careful selections of representative examples is introduced. Finally, a novel discussion is presented, wherein the issues covered include suitability of SAR data models, understanding the multiplicative SAR data models, and two unique perspectives on constant false alarm rate (CFAR) detection: signal processing and pattern recognition. From a signal processing perspective, CFAR is shown to be a finite impulse response band-pass filter. From a statistical pattern recognition perspective, CFAR is shown to be a suboptimal one-class classifier: a Euclidean distance classifier and a quadratic discriminant with a missing term for one-parameter and two-parameter CFAR, respectively. We make a contribution toward enabling an objective design and implementation for target detection in SAR imagery. © *The Authors. Published by SPIE under a Creative Commons Attribution 3.0 Unported License. Distribution or reproduction of this work in whole or in part requires full attribution of the original publication, including its DOI.* [DOI: 10.1117/1.JRS.7.071598]

## 1 Introduction

Synthetic aperture radar (SAR) offers distinctive active remote sensing capabilities for both military and civilian applications. Target, clutter, and noise are three terms of military origins associated with automatic target recognition (ATR), and their definition depends on the application of interest. In the case of SAR imagery, target refers to the object(s) of interest in the imaged scene. Clutter refers to manmade (building, vehicles, etc.) and/or natural objects (trees, topological features, etc.) that tend to dominate the imaged scene. Noise refers to imperfections in the SAR image which are a result of electronic noise in the SAR sensor, as well as computational inaccuracies introduced by the SAR signal processor. The general structure of an end-to-end ATR system for SAR imagery (SAR-ATR), as reported in the literature, is depicted in Fig. 1. To account for the prohibitive amounts of processing pertaining to the input SAR imagery, the strategy is to divide and conquer. Accordingly, the SAR-ATR processing is split into three distinctive stages: detector (also known as prescreener), low-level classifier (LLC, also known as discriminator), and high-level classifier (HLC).[1–9] The first two stages together are commonly known as the focus-of-attention module. While this is the most common structure reported in the literature, it should be highlighted that (theoretically) there is no restriction on the number of stages.

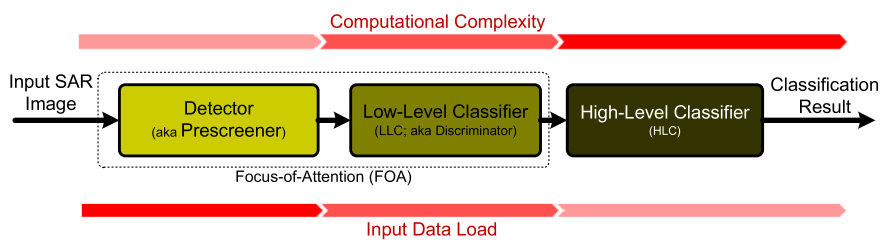

**Fig. 1** General structure for an end-to-end SAR-ATR system.

As depicted in Fig. 1, the input SAR image creates an extremely high computational load due to its high resolution and/or the presence of various clutter types and objects. As the SAR data progresses throughout the SAR-ATR processing chain, its load is reduced. The HLC stage deals with SAR data that has relatively lower computational load. To the contrary, the computational complexity of the SAR-ATR chain increases as the SAR data progresses from the *front-end* stage toward the *back-end* stage.

Detection is the front-end stage in any SAR-ATR processing chain. The detector interfaces with the input SAR image to identify all regions of interest (ROIs), thus ROIs can be passed-in to the LLC stage for further analysis. One may think of the detector as a dimensionality reduction scheme that properly reduces the dimensionality of the SAR data. The detector should be designed to balance the tradeoff between computational complexity, detection efficacy, and outlier rejection. On the one hand, it is required that the detector is relatively computationally simple, thus it can operate in real-time or near-real-time. On the other hand, it is required that the detector enjoys a low probability of false alarm (PFA), and a high probability of detection (PD). Indeed, these often conflicting factors distinguish one detector from another.

There are numerous strategies for implementing the detector. This is evident in the overwhelming number of research articles published on the topic in the open literature. Different researchers tend to approach the topic from various perspectives. This makes it even more challenging and time consuming to relate the various research findings and to grasp the relationship between these various approaches. This shows a dire need for a survey that offers an umbrella under which various research activities can be broadly probed and taxonomized. This is precisely the goal of this paper.

In this paper, we restrict our attention to *single-channel* SAR imagery (i.e., single polarization). This is because the development of our survey is motivated by our endeavor to develop SAR-ATR algorithms for Spotlight mode Radarsat-2 data, which is a single channel.[10] However, many of the topics described in this survey are either applicable or extendable to multichannel SAR imagery. For readers interested in multichannel SAR image processing, please refer to pertinent references.[11–17]

The remainder of this paper is organized as follows. In Sec. 2, taxonomy of the detection methods is introduced. Primarily, three taxa are proposed: single-feature-based, multifeature-based, and expert-system-oriented. In Sec. 3, the various methods for implementing the detection module are comprehensively surveyed. First, a classification methodology for the various strategies under each taxon is introduced. Next, a description of the different sub-taxa is elaborated. The description commences with the Bayesian approach being the optimal approach, and it ends with the various detection strategies that fall under the expert-system-oriented taxon. Representative examples pertinent to SAR imagery are carefully chosen, and relevant comments are pinpointed throughout this section. The issues approached include proper choice of the size of the constant false alarm rate (CFAR) sliding window, CFAR as a suboptimal one-class classifier, CFAR loss, and cell averaging CFAR (CA-CFAR) as the baseline algorithm for comparison with other CFAR techniques. In Sec. 4, a compact tabular comparison between the surveyed methods is offered. In Sec. 5, a novel discussion is presented. The issues covered under the discussion include suitability of SAR data models, understanding the multiplicative SAR data models, and two unique perspectives on CFAR detection: a signal processing perspective and a statistical pattern recognition perspective. In Sec. 6, the paper is concluded.

## 2 Taxonomy of the Detection Methods

The detection module takes the entire SAR image and identifies the ROIs. Ultimately, the detected regions in the image are passed-in to the next stage in the SAR-ATR chain for further analysis. The goodness of any detection module is typically judged based upon three aspects of significance: computational complexity, PD, and false alarm rate (also known as PFA). The detection module should enjoy a low computational complexity such that it operates in real-time or near-real-time. This is in contrast to the succeeding stages in the SAR-ATR chain, which are relatively more computationally expensive. Further, a good detection module should provide a means to refine detections, reduce clutter false alarms, and pass ROIs; thus the detection method enjoys a reasonable PFA and acceptable PD.

We broadly taxonomize the detection algorithms reported in the open literature into three major taxa: single-feature-based, multifeature-based, and expert-system-oriented. This taxonomy is depicted in Fig. 2.

The single-feature-based taxon bases the detection in the SAR image on a single feature; typically the brightness in the pixel intensity commonly known as the radar cross-section (RCS). Various methods in the literature fall under this taxon. The single-feature-based approach is placed at the base of the pyramid in Fig. 2 because it is the most common and widely used in the literature. Further, the single-feature-based approach is the building block for the other two taxa.

The multifeature-based taxon bases the detection decision on a fusion of two or more features extracted from the input SAR image. Besides RCS, additional features that can be inferred and fused include multiresolution RCS and fractal dimension. Obviously, this taxon builds on the previous taxon and is expected to provide relatively improved detection performance, along with fewer false alarms. Multiple methods in the literature fall under this taxon.

Finally, the expert-system-oriented taxon is the most sophisticated. It extends the two aforementioned taxa and utilizes a multistage (two or more stages) artificial intelligence (AI) approach that bases the detection process in the SAR image on exploitation of prior knowledge about the imaged scene, clutter, and/or target(s). Prior knowledge is exploited through various means such as image segmentation, scene maps, previously gathered data, etc.

As the sophistication of the detection taxon increases, the complexity-performance trade-off arises. Caution should be taken when opting for a certain approach in order to balance this tradeoff carefully.

## 3 Taxa, Methods, and Selected Examples

Based on the aforementioned taxonomy, we broadly classify the various detection schemes and relevant methods reported in the literature in Fig. 3.

Primarily, under the single-feature-based taxon, the sliding window CFAR (CFAR-based) sub-taxon is the most popular. The various CFAR methods can be viewed through three perspectives. First, based on the specifications of the sliding window, there is fixed-size versus adaptive, as well as rectangle-shaped versus nonrectangle-shaped. Second, based on the strategy used to implement the CFAR technique, there are various strategies, including cell-averaging

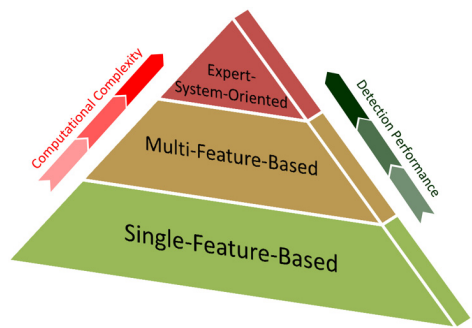

**Fig. 2** Major taxa for implementing the detection module.

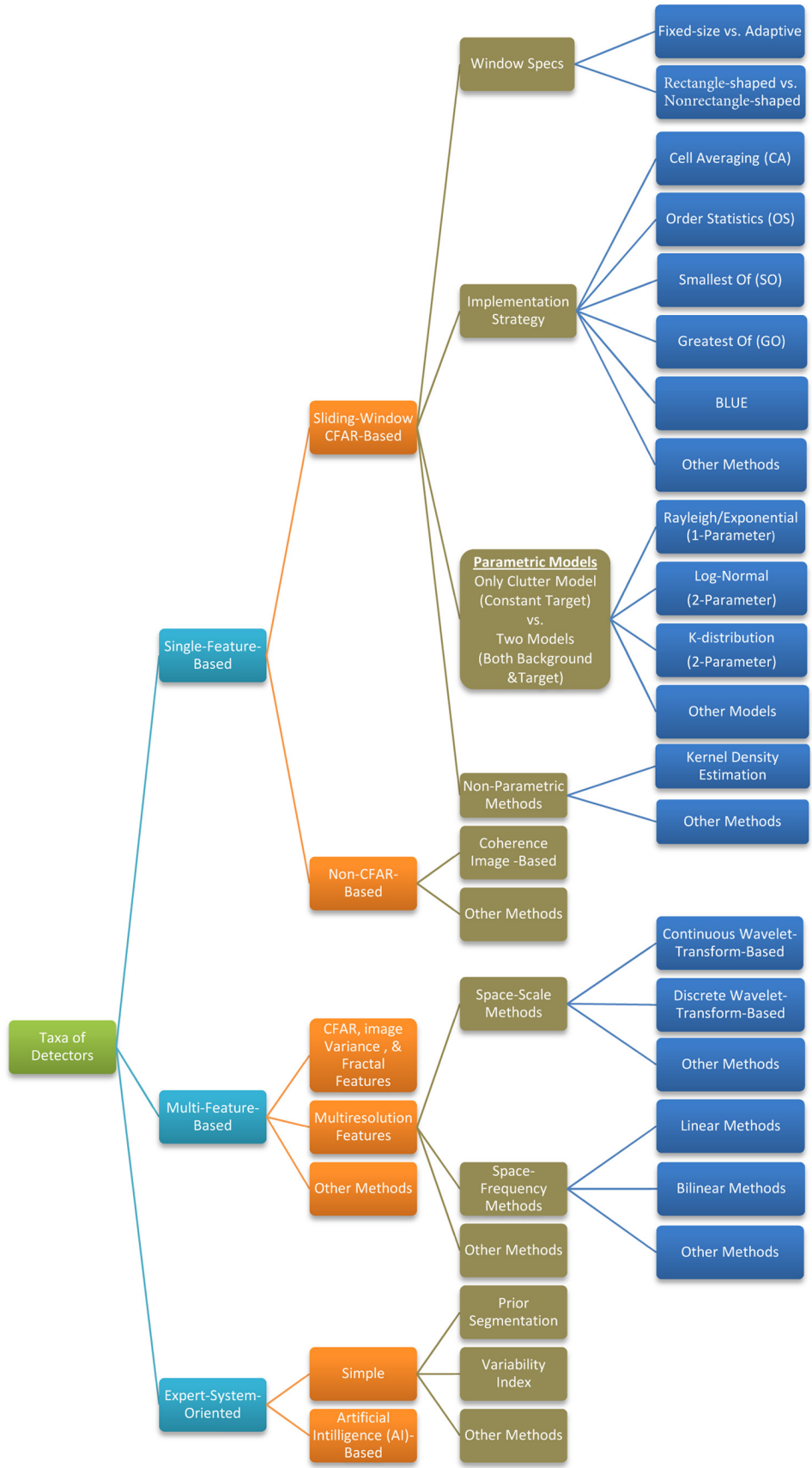

**Fig. 3** The three major target detection taxa and multiple sub-taxa and methods under each of them.

CFAR (CA-CFAR), smallest of CA-CFAR (SOCA-CFAR), greatest of CA-CFAR (GOCA-CFAR), and order statistics CFAR (OS-CFAR). Third, based on the method used to estimate the threshold (for a desired PFA) in the boundary ring and/or the approach utilized for estimating the target signature (for a desired PD), two subclasses emerge: parametric and nonparametric. Under the parametric subclass, two approaches are recognized: only background modeling and background and target modeling. A choice of the parametric model that best represents the SAR data in use has to be made among various parametric models. Unlike the parametric approach, the nonparametric approach does not assume any form for the background/target model(s). Rather, it directly infers an approximate model from the training data. One such method to perform the model inference is the kernel-density-estimation (KDE) method.

Less popular non-CFAR-based methods, such as those that rely on a coherence image, represent the other sub-taxon of single-feature-based methods. The single-feature-based taxon has the limitation that it bases the detection decision solely on RCS, and thus it can become overwhelmed in regions in the SAR image where there is heterogeneous clutter and/or a high density of targets.

Methods under the multifeature-based taxon try to circumvent this drawback by basing the detection decision on a fusion of two or more features. Obviously, this taxon can utilize a suitable method among those presented under the single-feature-based taxon and incorporate additional features besides RCS, such as multiresolution RCS analysis, fractal dimension, etc. Multiresolution methods can be either space-scale-based or space-frequency-based. Prime examples of methods that utilize space-scale features are those based on the wavelet transform, including the discrete wavelet transform (DWT), and the continuous wavelet transform (CWT). Prime examples of methods that utilize space-frequency features include linear space-frequency methods such as the Gabor transform and the S-transform, along with bilinear (also known as quadratic) space-frequency methods such as Cohen's class distributions (Wigner distribution, Wigner-Ville distribution, pseudo-Wigner-Ville distribution, etc.).

Finally, a more robust taxon is the expert-system-oriented approach, which incorporates intelligence into the process to guide the decision making. In its simplest form, detection decisions can be guided by a structure map of the imaged scene generated from properly segmenting the SAR image. Further, methods of AI can be appropriately integrated to achieve near-optimal context utilization.

Next we review the various methods introduced above under each taxon. Further, representative examples pertaining to SAR imagery under each method are carefully chosen and presented.

## 3.1 Single-Feature-Based Taxon

Single-feature-based detection algorithms base their search for target detection in the SAR image on a single feature. CFAR is the most popular single-feature-based detection algorithm. Despite the many variations of CFAR under this category, they are considered single-feature-based because they base the search for ROIs on RCS alone. Indeed, as it is evident from the numerous works published in the literature, CA-CFAR is the baseline approach for target detection in SAR imagery. To understand the limitations of the single-feature-based-CFAR approach, it is important to review its underlying assumptions.

An optimal detector (theoretically speaking) should utilize the Bayesian approach, which for a zero-one cost, reduces to the maximum *a posteriori* (MAP) criterion[18] as

$$\Lambda_{\mathrm{MAP}}(\mathbf{x}) = \frac{P(\omega_T|\mathbf{x})}{P(\omega_B|\mathbf{x})} \gtrless_{\omega_T}^{\omega_B} 1, \tag{1}$$

where $\omega_T$ is the target-class, $\omega_B$ is the background or clutter-class, and $P(\omega_T|x)$ and $P(\omega_B|x)$ are the posteriors of the target-class and the background-class, respectively.

This is simply a binary (i.e., two-class; dichotomizer) classification problem where $\mathbf{x}$ is a feature vector that represents the pixel values, typically obtained from the boundary ring in a sliding window with suitable guard cells centered on the ROI. This window is typically called a CFAR stencil and is depicted in Fig. 4 for a size of $9 \times 9$ pixels. The boundary ring is shown in bright green with suitable pixel labels.

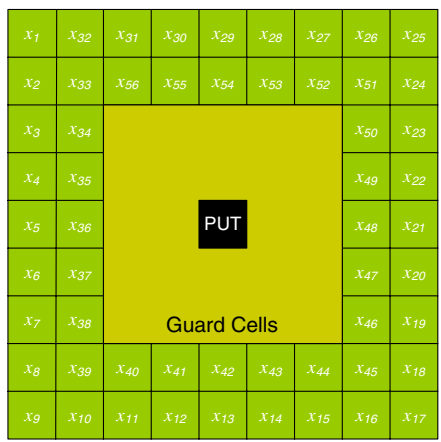

**Fig. 4** CFAR stencil.

Note that the choice of this stencil size here is for illustration purposes only. Proper choice of the stencil size will depend on the type of SAR image and the target size. More precisely, the size (and shape) of the guard ring will be governed by the geometrical size of the target.[19] However, it should be noted that choosing the stencil dimension using only the prior knowledge of target size yields a detection loss (i.e., CFAR loss that leads to a suboptimal performance), because the backscatter of the target in SAR imagery is dependent on the operating conditions and is weakly linked with the target's geometric shape.[20] Despite these challenges, it is recommended[21,22] that the target window size (i.e., pixels under test, PUTs) should be about the size of the smallest object that one wishes to detect, the guard ring window size should be about the size of the largest object, and the boundary ring window size should be large enough to estimate the local clutter statistics accurately.

Posterior probabilities can be expressed by the Bayes rule as

$$P(\omega_T|\boldsymbol{x}) = \frac{p(\boldsymbol{x}|\omega_T)P(\omega_T)}{p(\boldsymbol{x})}, \quad \text{and} \quad P(\omega_B|\boldsymbol{x}) = \frac{p(\boldsymbol{x}|\omega_B)P(\omega_B)}{p(\boldsymbol{x})}, \tag{2}$$

where $p(\boldsymbol{x}|\omega_T)$, and $p(\boldsymbol{x}|\omega_B)$ are the probability density functions (PDFs) or the likelihoods of the target-class and the background-class, respectively; $P(\omega_T)$, and $P(\omega_B)$ are the priors of the target-class and the background-class, respectively; And $p(\boldsymbol{x})$ is the evidence, which is typically ignored in the MAP criterion above, because it is identical for both classes.

Accordingly, the MAP criterion can be expressed as a likelihood ratio test (LRT) as

$$\Lambda_{\text{LRT}}(\boldsymbol{x}) = \frac{p(\boldsymbol{x}|\omega_T)}{p(\boldsymbol{x}|\omega_B)} \underset{\omega_T}{\overset{\omega_B}{\lessgtr}} \frac{P(\omega_B)}{P(\omega_T)}. \tag{3}$$

The major problem with this expression is that, in practice, we do not have knowledge on the class priors. If we assume equal priors for the target-class and the background-class, the LRT test reduces to the maximum likelihood (ML) test. However, in practice, these two priors are not equal, and the ML test should be avoided. Typically, the Neymann-Pearson (NP) criterion is adopted where the LRT test reduces to

$$\Lambda_{\text{NP}}(\mathbf{x}) = \frac{p(\mathbf{x}|\omega_T)}{p(\mathbf{x}|\omega_B)} \underset{\omega_T}{\overset{\omega_B}{\lessgtr}} \alpha, \tag{4}$$

where $\alpha$ is a detection threshold typically referred to as the threshold scaling factor for reasons that will become apparent.

Equation (4) is the main formula motivating the design for the CFAR algorithm and its variants. Indeed, many popular CFAR algorithms in the literature assume that only the background-class is characterized without characterizing the target-class, and thus they adopt a suboptimal anomaly detection (AD). This converts the optimal binary classification problem into a one-class classification problem[23] as

$$\Lambda_{\mathrm{AD}}(\mathbf{x}) = \; \mathrm{p}(\mathbf{x}|\omega_B) \underset{\omega_B}{\overset{\omega_T}{\lessgtr}} \alpha. \tag{5}$$

However, besides modeling the background, some other CFAR algorithms do consider a PDF model for the target-class. It is preferable to adopt such a target model, even if it merely represents a broad PDF (i.e., a weak prior assumption) rather than doing an AD[24,25] as shown in Eq. (5).

In the former approach (i.e., the AD approach), the CFAR algorithm assigns PUTs to the background if it finds that the PUTs are consistent with the background distribution; otherwise, the PUTs are labeled as detected. For a desired PFA, the scaling factor $\alpha$ is adaptively estimated throughout the image from the boundary ring in the sliding window focused on the ROI as

$$\mathrm{PFA} = \int_{\alpha}^{\infty} p(\mathbf{x}|\omega_B)\,\mathrm{dx}. \tag{6}$$

The latter approach (i.e., the one that models both the background-class and the target-class) utilizes the NP criterion as explained in Eq. (4), and the PD, for a certain target model of interest, $p(\boldsymbol{x}|\omega_T)$, is given by

$$\mathrm{PD} = \int_{\alpha}^{\infty} p(\boldsymbol{x}|\omega_T)\,\mathrm{dx}. \tag{7}$$

### 3.1.1 CFAR-based methods

Be it a fixed-size or adaptive sliding window, the various CFAR methods can be viewed through two perspectives intermingled. The first is based on the method used to estimate the threshold scaling factor (for a desired PFA) in the boundary ring and/or the approach utilized for estimating the target signature (for a desired PD). There are two strategies: parametric CFAR and nonparametric CFAR. The second is based on the method used to implement the CFAR technique. There are various strategies, including CA-CFAR, SOCA-CFAR, GOCA-CFAR, and OS-CFAR. Thus, any CFAR detector can be viewed as a combination of these two perspectives: one strategy pertinent to the estimation of the threshold scaling factor and one strategy for implementing the CFAR technique. To visualize the interrelation between the various methods and strategies, refer to Fig. 3.

Under this section, we review parametric CFAR and nonparametric CFAR. Under parametric CFAR, we review one-parameter CFAR and two-parameter CFAR. Under one-parameter CFAR, we review various implementation strategies, including CA-CFAR, SOCA-CFAR, GOCA-CFAR, and OS-CFAR. Under two-parameter CFAR, we discuss the most common implementation strategy. We then briefly discuss CFAR loss. Then an interesting remark that addresses an important issue pertaining to CFAR usage is presented. Finally, the topic of nonparametric CFAR is briefly approached.

*Parametric CFAR.* Parametric CFAR methods can be classified into two classes: methods based only on background modeling (i.e., AD) and methods based on both background and target modeling. We briefly review these two methods here.

All the parametric CFAR algorithms that perform AD (i.e., only models the background clutter) have one thing in common. They all assume that the background clutter can be roughly modeled by a certain probability distribution, i.e., $p(\boldsymbol{x}|\omega_B)$. Then, to perform the CFAR detection, they estimate the model distribution parameters from the boundary ring in the CFAR stencil. This PDF model is used to estimate the threshold scaling factor $\alpha$, for a desired CFAR (i.e., PFA), as the focusing window is systematically slid over the SAR image. However, variant classes of CFAR algorithms primarily differ in two aspects.

First, there is the probability distribution chosen, i.e., $p(\boldsymbol{x}|\omega_B)$, to model the background clutter. For example, some CFAR algorithms assume a homogeneous clutter and model the background clutter with an exponential distribution (i.e., for SAR image in the power-domain; magnitude-squared), or a Rayleigh distribution (i.e., for SAR image in the magnitude-domain). This class of distribution models is characterized by one-parameter (i.e., the mean) and thus is referred to in the literature as a one-parameter CFAR. Other CFAR algorithms model the background clutter in the SAR image with the more realistic but more complex Weibull distribution,[26,27] $K$-distribution,[28] alpha-stable distribution,[29,30] or beta-prime ($\beta'$) distribution,[31] among other models. This class of distribution models is characterized by two parameters (mean and variance, scale and shape parameters, etc.), and because of this, the CFAR algorithm is referred to in the literature as a two-parameter CFAR.

Second, there is the method used to estimate the model parameters pertaining to the detection Threshold from the boundary ring. For example, there is CA-CFAR, GOCA-CFAR, SOCA-CFAR, OS-CFAR,[32–34] and best linear unbiased estimator CFAR (BLUE-CFAR),[26,27] among others.

The CFAR algorithms that model both the background and the target perform procedures similar to those mentioned above. However, besides estimating the background parameters in the focusing window, they also estimate the target model parameters. Thus, the detection threshold in the sliding window is determined based on the NP criterion as shown in Eq. (4). Gan and Wang[35] and Rong-Bing and Jian-Guo[36] offer examples on this approach.

*One-parameter CFAR.* Under this section, we briefly review various one-parameter CFAR implementation strategies, including CA-CFAR, SOCA-CFAR, GOCA-CFAR, and OS-CFAR. Obviously, these same strategies are also utilized for implementing two-parameter CFAR. Accordingly, a proper understanding of these implementation strategies paves the way for better understanding two-parameter CFAR.

CA-CFAR was the first CFAR test proposed in 1968 by Fin and Johnson.[37–39] The adaptive Threshold is comprised of two parts. One is estimated from the boundary ring called $Z$, and the other is found from the corresponding PDF distribution for a desired PFA. This explains why $\alpha$ is referred to as the threshold scaling factor. Thus, Threshold is given by

$$\text{Threshold} = \alpha Z. \tag{8}$$

The underpinning assumption is that the clutter in the in-phase ($I$) and quadrature ($Q$) channels of the radar (particularly speckle, as no interfering targets are assumed) is Gaussian. Note that the focused SAR data output from the SAR processor (i.e., level-1 processed) is a complex-valued image of the form, $I + jQ$. The CA-CFAR detection can be performed on the magnitude SAR image (i.e., $A = \sqrt{I^2 + Q^2}$), the power SAR image (i.e., $P = A^2$), or the log-domain image (i.e., $L = 10 \log A^2$). It should be highlighted that a CFAR detector that deals with a magnitude SAR image is commonly known in the literature as an envelope (i.e., magnitude) detector or linear detector. A CFAR detector designed to handle a power SAR image is known as a square-law (i.e., power) detector. Finally, a CFAR detector designed for a log-domain SAR image is known as a log detector. Note that this terminology is also applicable to the two-parameter CFAR method. Accordingly, the clutter in the SAR image will be Rayleigh or exponential distributed depending on whether the SAR image is magnitude or power, respectively. Obviously, both the exponential distribution and Rayleigh distribution are completely characterized by one parameter (i.e., mean), thus CA-CFAR is also known as one-parameter CFAR.

Assuming a power image, and assuming the clutter is independent and identically distributed (iid), and that the PDF of a pixel $x_i$ is exponential distributed, whereby both the $I$ and $Q$ channels with power $Z/2$ in each channel (total $Z$), as follows:

$$p(\boldsymbol{x}_i | \omega_\text{B}) = \frac{1}{Z} e^{-x_i/Z}. \tag{9}$$

Thus, for $N$ reference pixels in the boundary ring (see Fig. 4),

$$\Gamma = p(\boldsymbol{x}|\omega_B) = \frac{1}{Z^N} \prod_{i=1}^{N} e^{-x_i/Z} = \frac{1}{Z^N} e^{-\frac{\sum_{i=1}^{N} x_i}{Z}}. \tag{10}$$

From this, $Z$ is approximated from the log maximum likelihood estimate (log-MLE) as

$$\frac{d}{dZ} \ln \Gamma = \frac{d}{dZ} \left( -\frac{\sum_{i=1}^{N} x_i}{Z} - N \ln Z \right) = 0. \tag{11}$$

Thus,

$$\hat{\mu}_B = \hat{Z} = \frac{\sum_{i=1}^{N} x_i}{N}. \tag{12}$$

Note that $\hat{Z}$ is simply the maximum likelihood estimate (MLE) of the arithmetic mean of the pixels in the boundary ring in the CFAR stencil. Accordingly, from Eq. (8),

$$\text{Threshold} = \alpha Z \approx \alpha \hat{\mu}_B = \alpha \frac{\sum_{i=1}^{N} x_i}{N}. \tag{13}$$

The scaling factor is estimated based on Eq. (6), and for this case is given[40] by

$$\alpha = N(\text{PFA}^{\frac{-1}{N}} - 1). \tag{14}$$

Since PFA does not depend on the average power $Z$ in the reference cell, the algorithm is CFAR under the assumptions mentioned earlier in this section.[40]

In reference to Fig. 4, CA-CFAR computes the arithmetic average of the pixels in the boundary ring and then compares it with the PUT. The decision to conclude either that a PUT is detected (i.e., target-class) or not detected (i.e., background-class) is contingent upon the threshold scaling factor $\alpha$, as illustrated below. The SAR image is in the (noncomplex) power domain:

$$\frac{X_{\text{PUT}}}{\hat{\mu}_B} \underset{\omega_T}{\overset{\omega_B}{\lessgtr}} \alpha, \tag{15}$$

where $\hat{\mu}_B$ is calculated based on Eq. (12) whereby $N$ is the total number of pixels in the boundary ring ($N = 56$ in Fig. 4), and $x_i$ is a pixel value in the boundary ring, and $X_{\text{PUT}}$ is the PUT.

Note that if a log-detector CFAR is used (i.e., the SAR image is assumed to be in the log-domain), one simply takes the logarithm for both sides in Eq. (15). Accordingly, the log-detector one-parameter CFAR is governed by

$$X_{\text{PUT}_{\log}} - \hat{\mu}_{B \log} \underset{\omega_T}{\overset{\omega_B}{\lessgtr}} \alpha_{\log}. \tag{16}$$

The subscript log is inserted in the above inequality to differentiate it from Eq. (15) and to emphasize the fact that all the terms in Eq. (16) are calculated based on the SAR image in the log-domain.

Additionally, note that it is possible to consider more than one pixel for the PUT in the inequalities above. However, in this case, $X_{\text{PUT}}$ should be replaced with the ML estimate of the arithmetic average pertaining to the $M$ PUTs given by

$$\bar{X}_{\text{PUT}} = \frac{\sum_{i=1}^{M} x_i}{M}. \tag{17}$$

CA-CFAR relies on two major assumptions that pose limitations on its performance. First, the targets are assumed to be isolated by at least the stencil size, so that there is at most one target in the stencil at one time (i.e., no interfering targets are assumed). Second, the reference pixels in the boundary ring are assumed to be iid distributed and have a PDF similar to that in the PUT (i.e., homogeneous clutter). Obviously, these assumptions do not hold in many real-world scenarios. Thus, when CA-CFAR is used under circumstances different from the design assumptions, its performance suffers a significant detection loss[40] (also known as CFAR loss).

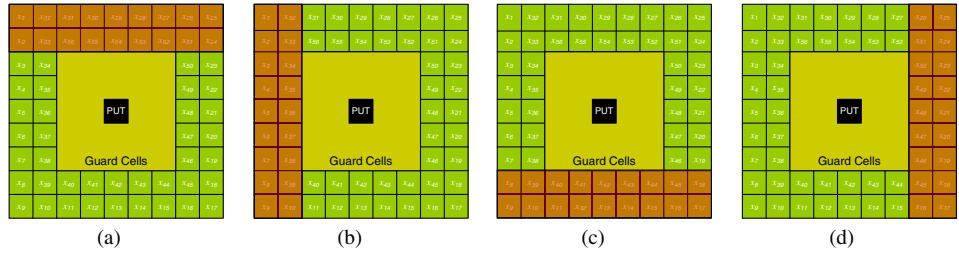

**Fig. 5** Four different strategies for defining the leading (a), (b); and lagging (c), (d) windows. Pixels shaded in orange illustrate a leading or lagging window in relation to the PUT.

SOCA-CFAR and GOCA-CFAR are variants of CA-CFAR that split the boundary ring in the sliding stencil into separate leading and lagging windows (indeed, there are four such windows, one on each side of the PUT) that are used to estimate separate statistics.[41] This is depicted in Fig. 5.

In SOCA-CFAR, the smallest of the four mean estimates is used to perform the test, while in GOCA-CFAR, the greatest of the four mean estimates is used. Assume there are a total of $N$ pixels in each window. Thus, there are four mean estimates:

$$\text{mean}_{\text{top}} = \frac{\sum_{i=1}^{N} x_{i,\text{top}}}{N}, \qquad \text{mean}_{\text{left}} = \frac{\sum_{i=1}^{N} x_{i,\text{left}}}{N},$$

$$\text{mean}_{\text{bottom}} = \frac{\sum_{i=1}^{N} x_{i,\text{bottom}}}{N}, \quad \text{and} \quad \text{mean}_{\text{right}} = \frac{\sum_{i=1}^{N} x_{i,\text{right}}}{N}, \tag{18}$$

where $\text{mean}_{\text{top}}$, $\text{mean}_{\text{left}}$, $\text{mean}_{\text{bottom}}$, and $\text{mean}_{\text{right}}$ are the arithmetic average estimates for the top, left, bottom, and right windows, respectively, and $x_i$ is the corresponding pixel value inside each window.

Accordingly, for SOCA-CFAR, the detection decision for power-domain (i.e., square-law) detection is

$$\frac{X_{\text{PUT}}}{\min\{\text{mean}_{\text{top}}, \text{mean}_{\text{left}}, \text{mean}_{\text{bottom}}, \text{mean}_{\text{right}}\}} \underset{\omega_T}{\overset{\omega_B}{\gtrless}} \alpha. \tag{19}$$

Similarly, for GOCA-CFAR, the power-domain detection decision is

$$\frac{X_{\text{PUT}}}{\max\{\text{mean}_{\text{top}}, \text{mean}_{\text{left}}, \text{mean}_{\text{bottom}}, \text{mean}_{\text{right}}\}} \underset{\omega_T}{\overset{\omega_B}{\gtrless}} \alpha. \tag{20}$$

Note that, similar to CA-CFAR, the threshold scaling factor $\alpha$ is determined based on the desired PFA and the (Gaussian) distribution used to model the clutter in the relevant reference window. Also note that the relationship between $\alpha$ and PFA will be different from that given in Eq. (14). For derivation of the relationship for GOCA-CFAR, the reader is referred to Chap. 7, pages 365, and 367, respectively, in Ref. 40.

SOCA-CFAR is designed to handle strong clutter returns in the boundary ring, but it is susceptible to clutter edges. On the other hand, GOCA-CFAR is capable of performing better than CA-CFAR and SOCA-CFAR at clutter edges, but its performance degrades when strong returns appear in the boundary ring. Further, compared to CA-CFAR, both SOCA and GOCA suffer from additional CFAR loss, due to the consideration of only a part of the boundary ring.

OS-CFAR was first proposed by Rohling[42] to counter multiple targets in the CFAR stencil. OS-CFAR rank orders the pixels in the boundary ring according to their value. Assuming that there are N pixels in the boundary ring of Fig. 4, OS-CFAR orders them in an ascending order:

$$x_{(1)} \geq x_{(2)} \geq \ldots \geq x_{(N)}. \tag{21}$$

Further, the $Q$'th percentile is chosen in place of the average estimate in CA-CFAR. Thus the detection decision is based on

$$\frac{X_{\text{PUT}}}{x_{(Q)}} \lessgtr^{\omega_B}_{\omega_T} \alpha. \tag{22}$$

Similar to earlier CFAR methods, the threshold scaling factor $\alpha$ is estimated based on the clutter statistics in the boundary ring. In Rohling's original work,[42] $Q = 3/4$ is found to work best. However, in later work,[43,44] $Q = 4/5$ is found to give better results. Obviously, the best value of $Q$ is dependent on the type of the SAR data used. Thus, bearing these values in mind, it is desirable to empirically check the value of $Q$ that best suits the data in use. For derivation of the threshold scaling factor $\alpha$ for a desired PFA under the Gaussian assumption, the reader is referred to page 372 in Richards.[40]

In heterogeneous/nonhomogeneous clutter backgrounds, and for contiguous targets, OS-CFAR is known to outperform CA-CFAR.[38,42,45,46] However, the performance of OS-CFAR degrades during clutter transitions.[45] This motivated researchers to develop this method further to handle various scenarios. Although only tested on 1-D radar data (i.e., range profiles), a variant called switched order statistics (SWOS) CFAR is designed for detecting targets in non-homogenous clutter and/or multiple interfering target scenarios.[47] This algorithm builds on a relevant method known as selection and estimation[48] (SE) and is able to determine whether the cells in the boundary ring belong to homogeneous or nonhomogeneous clutter and thus adaptively adjust the detection threshold for a desired PFA. SWOS-CFAR is shown to outperform standard OS-CFAR[49] and switching CFAR (S-CFAR). A generalized version of S-CFAR termed GS-CFAR is proposed, wherein it is shown that GS-CFAR yields some improvement in the detection probability in the case of interfering targets and clutter edges.[50] Many additional variations of CFAR exist in Ref. 40.

Gandhi and Kassam[51] show that the performance of the CA-CFAR processor approaches that of the NP detector in that the limit of the number of pixels in the boundary ring approaches infinity under the condition that homogeneity is maintained in the boundary ring. Thus, CA-CFAR achieves asymptotically optimal performance under these (theoretical) assumptions. Accordingly, CA-CFAR can be used as the baseline algorithm for comparison with other CFAR techniques.

We summarize the four basic (square-law) CFAR detectors (i.e., CA-CFAR, SOCA-CFAR, GOCA-CFAR, and OS-CFAR) in Table 1.

*Two-parameter CFAR.* Unlike one-parameter distribution models discussed earlier, more realistic two-parameter distribution models characterize the clutter in the boundary ring of the CFAR stencil by two parameters (mean and variance, scale and shape, etc.). Examples of two-parameter distribution models include log-normal distribution and Weibull distribution. For high-resolution SAR imagery, compound two-parameter distributions such as $K$-distribution, $G^o$-distribution, and $\beta$'-distribution are typically used.

A conventional two-parameter CA-CFAR algorithm based on the log detector has the form[14,52,53]

**Table 1** Summary of the four basic square-law CFAR techniques.

| Method | Formula | Advantages | Disadvantages |
|---|---|---|---|
| CA-CFAR | $\dfrac{X_{\text{PUT}}}{\frac{1}{N}\sum_{i=1}^{N} x_i} \lessgtr^{\omega_B}_{\omega_T} \alpha$ | Optimal in homogeneous clutter (the baseline detector) | Susceptible to nonhomogeneous clutter in the boundary ring |
| SOCA-CFAR | $\dfrac{X_{\text{PUT}}}{\min\{.\}} \lessgtr^{\omega_B}_{\omega_T} \alpha$ | Designed to handle strong clutter returns in boundary ring | Susceptible to clutter edges in the boundary ring |
| GOCA-FAR | $\dfrac{X_{\text{PUT}}}{\max\{.\}} \lessgtr^{\omega_B}_{\omega_T} \alpha$ | Perform well on clutter edges | Susceptible to strong returns appear in the boundary ring |
| OS-CFAR | $\dfrac{X_{\text{PUT}}}{x_{(Q)}} \lessgtr^{\omega_B}_{\omega_T} \alpha$ | Perform well in heterogeneous/ nonhomogeneous clutter backgrounds | Susceptible to clutter transitions |

$$\frac{X_{\text{PUT log}} - \hat{\mu}_{B \text{ log}}}{\hat{\sigma}_{B \text{ log}}} \underset{\omega_T}{\overset{\omega_B}{\gtrless}} \alpha_{\text{log}}, \tag{23}$$

where $\alpha_{\text{log}}$ is the threshold scaling factor estimated for a desired PFA based on the relevant model distribution for the background clutter.

Accordingly, rearranging Eq. (23) yields

$$\hat{\text{Threshold}}_{\text{log}} = \hat{\mu}_{B \text{ log}} + \alpha_{\text{log}} \hat{\sigma}_{B \text{ log}}, \tag{24}$$

where $\hat{\mu}_{B \text{ log}}$ and $\hat{\sigma}_{B \text{ log}}$ are the ML estimates of the background mean and standard deviation, respectively, calculated from the log-domain SAR image, and given by

$$\hat{\mu}_{B \text{ log}} = \frac{\sum_{i=1}^{N} x_i}{N} \quad \text{and} \quad \hat{\sigma}_{B \text{ log}} = \sqrt{\frac{1}{N} \sum_{i=1}^{N} (x_i - \hat{\mu}_{B \text{ log}})^2}, \tag{25}$$

where $x_i$ is the pixel in the boundary ring, and $N$ is the total number of reference pixels in the boundary ring.

PUT is assumed to be a single pixel. If more than one pixel is assumed, then PUT is the ML estimate of the arithmetic mean such that

$$\bar{X}_{\text{PUT}} = \frac{\sum_{i=1}^{M} x_i}{M}, \tag{26}$$

where $M$ is the number of pixels in the PUT.

However, note that this will entail replacing[54] $\hat{\sigma}_{B \text{ log}}$ in Eqs. (23) and (24) with $\hat{\sigma}_{m \text{ log}}$:

$$\hat{\sigma}_{m \text{ log}} = \frac{\hat{\sigma}_{B \text{ log}}}{\sqrt{M}}. \tag{27}$$

An embedded underlying assumption in deriving Eqs. (23) and (24) is that the background clutter in the $I$ and $Q$ channels of the SAR image follows a Gaussian distribution, and thus the clutter in the magnitude-domain SAR image or the power-domain SAR image is Rayleigh or exponential distributed, respectively. This assumption entails converting the SAR image to the log-domain. In some other works, the background clutter in the SAR image is assumed to be log-normal, and thus the above equations are adopted verbatim without applying the logarithmic conversion to the SAR image.[19,21,22,54]

If we keep the assumption implicit in Eqs. (23) and (24) (i.e., magnitude image is Rayleigh distributed, and power intensity image is exponential distributed), the two-parameter CFAR applied to the image in the (nonlog) domain is given[55] by

$$\frac{\frac{X_{\text{PUT}}}{\hat{\mu}_B} - 1}{\hat{\sigma}_B} \underset{\omega_T}{\overset{\omega_B}{\gtrless}} \alpha. \tag{28}$$

If more than one pixel is considered in the PUT, then a procedure similar to Eqs. (26) and (27) should be applied. Note that various CFAR combinations described in the previous section (i.e., SOCA, GOCA, OS, etc.) can also be applied to two-parameter CFAR.

*CFAR loss.* As it is explained thus far, the CFAR approach aims at maintaining a CFAR by locally adapting the detection threshold to the background clutter in the SAR image. However, a detection loss, commonly referred to as CFAR loss, is the price paid for this threshold adaptation. This is due to the fact that, in real-world applications, the noise level in the boundary ring is not constant and/or the number of reference pixels used in the estimation is not large enough. Further, CFAR loss can be viewed as the required increase in the signal to noise ratio (SNR) in order to maintain the desired PD.[56] The value of this CFAR loss is dependent upon a number of factors, including CFAR method used (e.g., CA, GOCA, SOCA, etc.), number of pixels in the boundary ring, PFA, target statistics, clutter statistics, and noise. Antipov and Baldwinson[57] offer a notable work on this topic, though applied to 1-D radar data.

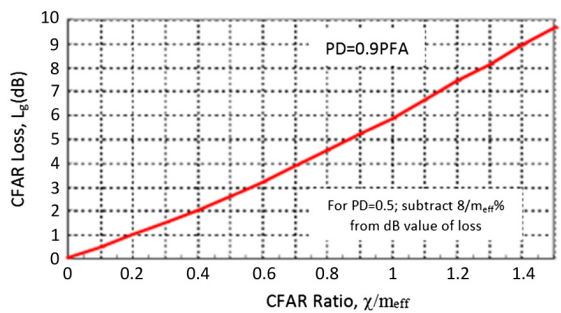

**Fig. 6** Universal curve of CFAR loss in single-hit detection for steady or Rayleigh target (See Refs. 56 and 57).

A universal curve for CFAR loss in a single-hit detection for steady or Rayleigh target[56] is depicted in Fig. 6. The parameter $\chi$ is given by

$$\chi = -\log \text{PFA}. \tag{29}$$

The CFAR ratio is given by

$$\text{CFAR Ratio} = \frac{\chi}{m_{\text{eff}}}, \tag{30}$$

where $m_{\text{eff}}$ is the effective number of reference pixels in the boundary ring, given by

$$m_{\text{eff}} = \frac{m + k}{1 + k}. \tag{31}$$

Values of $k$ for different CFAR detectors are provided in Table 2.

We now present a final word on the log-detector CFAR. Indeed, caution should be exercised when dealing with log-CFAR. As reported in Ref. 40 log-CFAR introduces an additional CFAR loss into the process. To circumvent this CFAR loss, the size of the CFAR stencil, more precisely the boundary ring in the stencil, needs to be increased by 65 percent[58] by following

$$N_{\text{log}} = 1.65 \, N - 0.65, \tag{32}$$

where $N$ is the number of pixels used for the nonlog CA-CFAR detector, and $N_{\text{log}}$ is the number of corresponding pixels required for the log CA-CFAR detector to circumvent the CFAR loss due to using the log-detector.

Indeed, despite this additional hurdle, log-detection CFAR is shown to be preferred over linear threshold CFAR processing for typical nonhomogeneous clutter conditions where background clutter surrounding the target is contaminated by other targets, bright clutter pixels, etc.[59,60]

**Table 2** Values of $k$ under different CFAR detectors (Ref. 56).

| CFAR Method | SAR image type | $k$ |
|---|---|---|
| CA-CFAR | Square-law detector | $k = 0$ |
| | Linear envelope detector | $k = 0.09$ |
| | Log detector | $k = 0.65$ |
| GOCA-CFAR | Square-law detector | $k = 0.37$ |
| | Linear envelope detector | $k = 0.5$ |
| | Log detector | $k = 1.26$ |

*Final remarks on parametric CFAR.* This section highlights the fact that the CFAR detectors described earlier that are originally designed under the Gaussian assumption (i.e., the background clutter distribution in the $I$ and $Q$ channels) are typically applied in the literature to nonGaussian clutter. Typically, a suitable clutter model (Weibull distribution, $K$-distribution, $\beta$'-distribution, etc.) is adopted, wherein the distribution model parameters are estimated from the boundary ring in the sliding stencil and are used to estimate the scaling factor ($\alpha$) for the desired PFA.

In Ref. 19, the two-parameter CFAR scheme is originally designed under Rayleigh/Gaussian assumptions (similar to the one presented earlier) and applied to nonGaussian data in high-resolution SAR images. In Armstrong and Griffiths,[61] the performance of two-parameter CA-CFAR, GOCA-CFAR, and OS-CFAR (originally designed under Gaussian assumption) is evaluated under $K$-distributed clutter. In Refs. 31 and 62 two-parameter CA-CFAR (originally designed based on Gaussian assumption) is applied to $\beta$'-distributed high-resolution SAR data. Additional examples are provided in Refs. 53 and 54. Obviously, such CFAR detection schemes are applied to clutter distributions that are different from those on which the CFAR detector was originally designed. Subsequently, the conclusion is reached that CFAR techniques prove to be effective regardless of this fact, which explains their popularity.[63] However, CFAR loss is the price paid for this process.[40]

*Nonparametric CFAR.* In the parametric CFAR methods explained in the preceding sections, the background clutter and/or the target pixels are assumed to obey a certain distribution, and the pixels in the reference window (i.e., boundary ring) are used to estimate the corresponding model parameters. Nonparametric CFAR algorithms do not assume any prior model for the background or the target. Rather, they use nonparametric methods[18,64] to directly infer the model from the SAR data. An example on a nonparametric CFAR using kernel density estimation (KDE) for inferring the background and the target models is proposed in Gao.[65]

### 3.1.2 *Nonrectangle-shaped-window methods*

While most detection techniques reported in the literature rely on a rectangle-shaped hollow stencil with a suitable size and a guard ring, there are some other methods that replace the rectangle-shaped window with another shape. One such example is provided in Ref. 66, where the rectangle-shaped CFAR window is replaced with a 2-D gamma kernel. The method utilizes a CFAR detection strategy and is referred to as gamma-CFAR. Similar to traditional rectangle-shaped-window CFAR, the CFAR threshold in gamma-CFAR is estimated from the radial pixel intensity information around a PUT. Further, gamma-CFAR has a free parameter which can be used to estimate the size of its region of support and its standard deviation required for the CFAR test. Although it is not implemented in Ref. 66 it is stated that this free parameter can be set adaptively. The gamma stencil is also applied in a non-CFAR fashion based on a quadratic Gamma discriminant (QGD) that has eight free parameters.[66] Test results show that QGD outperforms gamma-CFAR. Extension of QGD to an artificial neural network (ANN) based on a multilayer perceptron (MLP) is provided in Ref. 67.

### 3.1.3 *Non-CFAR methods*

Besides CFAR-based methods, there are other approaches that do not use CFAR. For example, in Ouchi et al.,[68] the detection is based on a coherence image produced from the multilook SAR image via cross correlation between two SAR images extracted by moving windows of a small size over the original image. The method can detect objects buried in a speckle. In Howard, Roberts, and Brankin,[69] the detection is based on genetic programming. Relevant work is provided in Conte, Lops, and Ricci,[70] although applied to 1-D radar data, where the detection is based on a generalized likelihood ratio test (GLRT) in a $K$-distributed clutter. The null hypothesis ($H_o$) represents the $K$-distributed clutter, and the alternative hypothesis ($H_1$) is modeled as being a compound of target signal and a $K$-distributed noise.

## 3.2 Multifeature-Based Taxon

All the aforementioned methods for target detection are single-feature-based in that they distinguish the target pixels from the background only on the basis of the level of pixel brightness (i.e., RCS). This poses a limitation on these methods, more significantly, in ROIs with heterogeneous clutter and/or contiguous targets.

Methods under the multifeature-based taxon try to circumvent this drawback by basing the detection decision on a fusion of two or more features. Obviously, this taxon can utilize a suitable method among those presented under the single-feature-based taxon and incorporate additional features besides RCS. Methods that fall under the multifeature-based taxon can be broadly classified into two major sub-taxa: those that utilize arbitrary user-chosen features and those that rely on systematic multiresolution analysis. Examples of arbitrary user-chosen features are provided in Refs. 9 and 71, wherein decisions on ROIs in the SAR image are based on a fusion of three multistage features extracted in parallel from the SAR image, namely CFAR features, variance features, and extended fractal (EF) features. Obviously, this approach is not purely CFAR. Another relevant example of the multifeature method is provided in Ref. 72. Further, in Subotic et al.,[73] parametric circular complex Gaussian models are utilized for both target and clutter. However, unlike the traditional CFAR approach, which works on a single resolution, the resolution of the SAR image in this approach is varied to produce multiple features. It is based on the conjecture that target signatures, when viewed as a function of resolution, are sufficiently different from clutter. This allows detection performance gains over single-resolution detection algorithms.

This perspective motivates the more systematic multiresolution analysis methods, which can be broadly classified into space-scale-based and space-frequency-based. Space-scale methods produce space-scale features based on the wavelet transform, including the DWT, and CWT. In Ref. 74 the detection is based on DWT that yields a spatial correlation of the sub-bands that attenuates the background noise and enhances the structured pattern of the target signature. A description of the more relevant detection strategies based on *CWT* can be found in Antoine et al.[75]

Prime examples of methods that utilize space-frequency features include linear space-frequency methods, such as the S-transform,[76] and bilinear (also known as quadratic) space-frequency methods, such as Cohen's class distributions[77,78] (Wigner distribution, Wigner-Ville distribution, pseudo-Wigner-Ville distribution, etc.).

## 3.3 Expert-System-Oriented Taxon

Expert-system-oriented target detection is a multistage (two or more stages) AI approach that bases the detection process on exploitation of prior knowledge about the imaged scene and/or target(s). Prior knowledge is captured via context utilization. In the broader sense, context here refers to all the available means that can help to convey information about the background clutter and/or target(s) of interest in the scene. Such means include image segmentation, scene maps, digital elevation model (DEM), previously gathered data, and geographical information system (GIS).

In its simplest form, context utilization in CFAR can be realized via unsupervised/semisupervised SAR image segmentation. Typically, prior to applying CFAR to the SAR image, the SAR image is segmented to extract an image structure map. Then, the conventional CFAR method of choice is aided with this map to enable it to adaptively select the suitable area over which the background statistics can be properly estimated. Further, smaller segments can be labeled as potential targets, while larger ones can be labeled as background. An example for utilizing annealed segmentation with one-parameter CFAR for SAR target detection is provided in McConnell and Oliver.[79]

A relevant work is termed variability index CFAR (VI-CFAR), although applied on 1-D range data.[80,81] Based on the estimations of the mean in the boundary ring of the stencil, VI-CFAR switches between one of the following three CFAR methods: CA-CFAR, SOCA-CFAR, and GOCA-CFAR. This approach tries to benefit from the strength of each CFAR method through deployment of the switching scheme.

Another relevant algorithm is reported in Gao et al.[62] The algorithm builds on the beta-prime CFAR ($\beta'$-CFAR) algorithm[31] reported earlier in this paper. A binary index map is created based on globally thresholding the input SAR image. The index map is comprised of zeros assigned to pixels in the input SAR image that are found to be less than a predetermined global threshold, and ones assigned to pixels found to be greater than the global threshold. Then, a sliding window stencil is placed over the image (i.e., pixel by pixel) where the parameters of the $G^o$-distribution (i.e., $G^o$-distribution reduces to $\beta'$-distribution for a single look SAR image; more on this is given in Sec. 5.2) are estimated from the pixel values in the boundary ring correspond to 0 in the index map. The size of the stencil is chosen based on the prior knowledge of the target size as described in Salazar.[31] Then, a local threshold in the stencil is determined based on the parameters estimated, and a decision on the PUT is taken. The window is then systematically slid to subsequent pixels in the image until the whole image is scanned. Further refinements on detections are achieved by placing a constraint on the size and allowable distance between detections.

More robust expert-system-oriented approaches utilize a mixture of multiple different CFAR experts (CA-CFAR, OS-CFAR, etc.), each of which is specialized to handle a suitable kind of clutter. The expert-system-oriented system uses available context information, extracted by one or more means (as explained earlier in this section), to assign the CFAR experts to suitable regions in the SAR image. Unfortunately, there is a lack of work published on this approach for SAR imagery. However, there are works published on 1-D radar data. One such example[82–84] was implemented by the U.S. Air Force Research Laboratory.

Rimbert and Bell[43,44] present another interesting work. It is motivated by the observation that, in homogeneous regions (i.e., locally stationary clutter), a larger reference window provides a clutter estimate that has a mean with smaller variance than a similar estimate based on a smaller reference window. Conversely, a smaller reference window provides a more reliable mean estimate in nonhomogeneous regions. Accordingly, a target detection scheme that adapts the reference window size and selects one of two CFAR detectors (i.e., CA-CFAR, and OS-CFAR) based on the type of the reference region is proposed. The proposed detection scheme checks the homogeneity of the reference region using a goodness-of-fit statistical test with an assumed parametric model for the clutter. One can think of this process as producing a structure map for the reference window (i.e., not for the whole image, as discussed in this section for the segmentation-based method). A CFAR method termed cell under test inclusive CFAR (CI-CFAR) is also introduced, which operates in a manner similar to OS-CFAR but combines the PUT with the clutter pixels in the reference window. Although the authors reported in one work[43] that CI-CFAR outperforms OS-CFAR, in another work,[44] the authors acknowledge that this conclusion is incorrect, due to the use of a simulation with errant detection threshold parameters. It should be noted that the results presented are not based on 2-D SAR data. Further, the analysis was based on the assumption that both the clutter and the target obey the central limit theorem (CLT) (i.e., Gaussian), which is a nonrealistic assumption for high-resolution SAR imagery. Additionally, reference stencils used are not hollow, and they do not consider any guard regions.

Finally, any relevant detection method that utilizes any form of intelligence/inference is of particular interest and to fit under this taxon.

## 4 Comparison

A comparison between selected examples, most of which are from amongst those cited in Sec. 3, on the detection module pertaining to SAR imagery is provided in Table 3. The choice of the examples was carefully made to cover the different methods under each taxon. Comparison aspects include SAR image type, feature(s), clutter/target type, clutter model (if applicable), and target model (if applicable). The comparison does not attempt to assess the algorithm performance based on reported PD and PFA, as this is infeasible, given the variant types of methods and data used, as well as the different sensor characteristics and operating conditions. Rather, the aim of this comparison is to show the major differences between the various methods and depict their applicability to certain scenarios.

**Table 3** Comparison between selected detection modules [$T$ = Taxon, ST = Sub-taxon].

| $T$ | ST | Refs. | Image type | Feature(s) | Clutter/target | Clutter model | Target model | Comments |
|---|---|---|---|---|---|---|---|---|
| Single-feature-based | CFAR-Based | | **Parametric methods: based only on background modeling** | | | | | |
| | | 85 | SIR-C/X SAR data | One/RCS | Sea/ship | Joint log-normal | N/A | Uses sliding window CA-CFAR. Tested on isolated targets in a homogeneous clutter. |
| | | 33 | HH/ airborne/ high resolution | One/RCS | Homogeneous/ glinting area targets | Exponential | N/A | Uses sliding window OS-CFAR. KDE is used to estimate PFA and PD. Logarithm of intensity is considered for calculations. |
| | | 26, 27 | Spotlight/ HH/ airborne/ high resolution | One/RCS | Land/extended objects | Weibull | N/A | Uses sliding window CFAR. Homogeneous clutter in local windows is assumed. Location-scale type (BLUE estimate)[96] is used for parameter estimation. |
| | | 28, 86 | Lincoln Lab/HH and HV | One/RCS | Land/bridge, and power-line tower | K-distribution | N/A | Uses sliding window CFAR. OS-CFAR and CA-CFAR are compared under Weibull and $K$-distribution clutters. |
| | | | **Parametric methods: based on background and target modeling** | | | | | |
| | | 35, 36 | High resolution (no info provided) | One/RCS | Land/vehicles | Log-normal | Log-normal | Uses sliding window-CFAR. (not AD). NP criterion is utilized. |
| | | | **Nonparametric methods: based on kernel density estimation (KDE)** | | | | | |
| | | 65 | Radarsat-1/space-borne/high resolution | One/RCS | Sea/ship (homogeneous clutter, and isolated targets) | Nonparametric. Estimated using KDE | N/A | Uses sliding window CFAR, for AD. Gaussian kernel is used for the KDE of the background model. |
| | Non-CFAR-based | | **Based on a coherence Image** | | | | | |
| | | 68 | Radarsat-1 | One/RCS (correlation-based) | Sea/ship (homogeneous clutter, and isolated targets) | N/A | N/A | A coherence image produced from the multilook image via cross correlation between two images extracted by moving windows of a small size over the original image. Can detect objects buried in speckle noise. |

**Table 3** (*Continued*).

| T | ST | Refs. | Image type | Feature(s) | Clutter/target | Clutter model | Target model | Comments |
|---|---|---|---|---|---|---|---|---|
| | | | | | **Based on genetic programming** | | | |
| | | 69 | Low resolution SAR imagery (ERS data) | One/RCS | Sea/ship (homogeneous clutter, and isolated targets) | N/A | N/A | Two-stage evolution strategy. |
| | | | | | **Nonrectangle/nonsquare-shaped-stencil methods** | | | |
| | CFAR | 66 | Fully polarimetric MIT high resolution SAR data. PWF is used to generate a single image. | One/RCS | Natural and cultural clutter (including a parking lot)/ vehicles | Gamma kernel with three parameter | N/A | The method is termed gamma-CFAR, and uses two 2-D gamma kernels to form the sliding stencil. PWF stands for polarimetric whitening filtering. |
| | Non-CFAR | 66 | Fully polarimetric MIT high resolution SAR data. PWF is used to generate a single image. | One/RCS | Natural and cultural clutter/ vehicles | QGD with eight parameters | N/A | The method is based on 2-D gamma kernels but replaces CFAR with QGD. QGD is shown to outperform gamma-CFAR. |
| Multi feature-based | CFAR Utilized | | | | **Based on multifeature fusion** | | | |
| | | 9, 71 | MSTAR, and TESAR emulated imagery. | Three features (RCS, variance, & extended fractals | Heterogeneous clutter/vehicles (isolated targets). | Gaussian | N/A | Features generation process is multistaged (four stages). Among other features, uses sliding window CFAR. Then detection of ROIs is based on fusion of the features. |
| | CFAR-Based | | | | **Based on multiresolution** | | | |
| | | 73 | Synthetic X-band SAR, and real SAR imagery | RCS at multi-resolutions | Land/vehicles (quite homogeneous clutter, and isolated target) | Zero-mean circular complex Gaussian | Circular Complex Gaussian | Sliding window-CFAR (Not AD). NP criterion is utilized. It is based on that target signatures, when viewed as a function of resolution, are sufficiently different from clutter. This allows detection performance gains over single-resolution detection algorithms. |

**Table 3** (*Continued*).

| T | ST | Refs. | Image type | Feature(s) | Clutter/target | Clutter model | Target model | Comments |
|---|---|---|---|---|---|---|---|---|
| | Non-CFAR-based | 74 | Synthetic (emulated), and Radarsat-1 | One/RCS | Sea/ship (homogeneous clutter, and isolated targets) | N/A | N/A | (Space-scale analysis) Application of DWT that yields a spatial correlation of the sub-bands attenuates background noise and enhances the structured pattern of ship signature. |
| Expert-System-Oriented | CFAR-Based | | | | **Based on segmentation** | | | |
| | | 79 | Simulated targets manually inserted in real and synthetic SAR images | Image structure, and RCS | Non-homogeneous clutter/ simulated targets on land inserted in the clutter. | Constant background is assumed. | N/A | Two-stage process including, segmentation and CFAR. CFAR criterion is based on AD. No model is assumed for the target. |
| | | | | | **Based on index matrix** | | | |
| | | 62 | Real SAR imagery, X-band, HH | Global thresholding, RCS, and target size | Non-homogeneous clutter/vehicle targets | $\beta'$-Distribution | N/A | The CFAR detection process is guided via a binary index matrix. Detections are refined via checking the distance between detection segments. The algorithm design utilizes the beta-prime CFAR algorithm reported in Ref. 31. |

## 5 Discussion

Obviously, target detection based on parametric modeling of SAR imagery is the most popular in the literature. This discussion focuses on parametric CFAR methods that utilize stochastic models for modeling the background clutter. Primarily, the issue of the suitability of these models to represent the SAR data is briefly discussed. This is followed by a concise summary of the popular multiplicative (also known as compound) SAR data models, conditions of their applicability, and the interrelation between them. Finally, our focus is shifted to parametric CFAR. We tame CFAR and present some novel discussions from two different perspectives: the signal processing perspective and the statistical pattern recognition perspective.

### 5.1 *On the Suitability of SAR Data Models*

In Table 3, there are various parametric clutter models used, including log-normal-distribution, $K$-distribution, and exponential distribution. Accordingly, in Table 4, we briefly summarize some of the major statistical distributions, along with the backscatter types in SAR images they are typically used to model.

**Table 4** Major statistical distributions and suitable modeling phenomenon.

| Distribution | Backscatter type/comments |
| --- | --- |
| Normal, and Rayleigh[87] | Homogenous (i.e., bare ground surfaces, dense forest canopies, snow covered ground). |
| Weibull, and log-normal[87] | Other clutter types such as sea surface. |
| Modified beta[88] | Different ice types. |
| $K$-distribution[89] | Models heterogeneous backgrounds. |
| | It offers a multiplicative model (compound distribution). |
| | $K$-distribution originally proposed in Ref. 90 for modeling microwave sea echoes. |
| | It then became popular for modeling multilook[91] and polarimetric SAR signature.[92] |
| | It has much poorer performance in extremely heterogonous clutter such as urban clutter.[93] |
| $G^o$-distribution[93] | Models extremely heterogeneous clutter background. Thus, it has better performance than $K$-distribution.[93,94] |
| | It is a compound distribution. |
| $G$-distribution[93] | It can model extremely heterogeneous clutter such as urban regions that $K$-distribution cannot.[93,94] |
| | It is a compound distribution. |
| | $K$ and $G^o$ distributions are special cases of this class. |

As is evident in Table 3 (and relevant works published in the literature), in many cases, similar distributions are being randomly chosen to model the clutter at various sensor characteristics, such as frequency, polarization, imaging mode (e.g., Spotlight, Stripmap, ScanSAR, etc.), and resolution. Obviously, the choice of a proper model for the clutter backscatter depends, not only on the clutter type, but also on these sensor characteristics. Several works[89,95,96] have noted that the suitability of a certain distribution to model a certain clutter depends on the data being used, as well as the corresponding sensor characteristics and operating conditions. Surprisingly, this issue is generally overlooked in many works published in the literature. In many such works, the justifications for choosing some model (e.g., $K$-distribution) is merely based on the assumption that the distribution model is found suitable for a certain clutter type (e.g., ocean clutter) in some published work, and thus it can be automatically adopted for a similar clutter. Indeed, such a conclusion can be misleading.

Accordingly, prior to randomly opting for a popular parametric distribution to model the clutter, one should consider validating the applicability of the distribution on the data using a suitable goodness-of-fit technique.[97,98] One such validation approach for SAR data is based on the Cramer-Von Mises (CVM) distance[97] as presented in di Bisceglie and Galdi.[26] The normally used Kolmogorov-Smirnov test is discarded, because the independency assumptions of its usage are violated, due to the dependency of the SAR data being non-Gaussian distributed and generally correlated.[97] Thus, the CVM method is used to measure the distributional distance between the design cumulative distribution function (CDF) model and the empirical CDF estimated from the available high-resolution SAR image. The design distribution that scores a minimum distance within some threshold is typically chosen. This procedure is demonstrated in di Bisceglie and Galdi[26] on Spotlight (9.6 GHz, HH polarized, geometric resolution of 0.4 m × 0.4 m) SAR data (with a specific clutter type) pertaining to the Rayleigh and Weibull distributions. On the SAR image used in di Bisceglie and Galdi,[26] the Weibull distribution is found to be a more suitable model for the clutter.

One final observation on this matter is that, though the CVM test was used in the work mentioned above, CVM is not the best technique when it comes to characterizing tailed

SAR data. Indeed, the $k$-sample Anderson-Darling test[99] offers a better procedure, as it places more weight on observations in the tail of the CDF distribution. Both the CVM test and Anderson-Darling test belong to the quadratic class of empirical distribution functions (EDF) for statistical tests.[99] Finally, information-theoretic approaches for characterizing the goodness-of-fit can be found in Ref. 100

### 5.2 *Understanding the Multiplicative SAR Data Models*

In any target detection scheme that depends on parametric modeling (e.g., the popular CFAR-based), the selection of an appropriate probability distribution to model the pixels in the SAR image (i.e., radar backscatter) is a must, because the thresholding operation in any such detector is dependent on the clutter distribution. In cases where the random scatterers in a resolution cell in the SAR image have sizes on the order of the wavelength of the radar signal, the total backscatter can be modeled as the sum of isolated returns in the cell.[101] This invokes the CLT, wherein the $I$ and $Q$ components of the total complex-valued backscatter can naturally be assumed to be normally distributed. This implies that the total backscattered amplitude and phase can be modeled as Rayleigh and uniform distributions, respectively. Thus, the power in each resolution cell is modeled as an exponential distribution. Conversely, in high-resolution SAR, the above mentioned assumptions are violated, because the number of random scatters in a resolution cell is not large, and thus the CLT cannot apply. This renders the clutter non-normally distributed, which motivates the need for a suitable model.

The multiplicative model (also known as the compound model) for SAR image formation has been popularly used in the literature to model the clutter background. The model is based on the hypothesis that the SAR image is formed from the product of a backscatter and speckle random processes as

$$Z = X \times Y, \tag{33}$$

where $X$ and $Y$ are two independent random variables that represent the backscatter and speckle, respectively. $X$ is often assumed to be a positive real variable, whereas $Y$ is either complex or positive real, depending on whether the image is in the complex or intensity/magnitude domains. The product in $Z$ models the observed SAR image.

Typically, for a single look (i.e., $n = 1$) SAR image, the complex speckle $Y$ is characterized as bivariate normal density for complex imagery, which reduces to the exponential distribution in the power or intensity domain. Further, for multilook imagery (i.e., $n > 1$), the two-parameter gamma distribution $\Gamma(\alpha, \lambda)$ characterizes the speckle in the power domain, and this reduces to the square root gamma distribution $\sqrt{\Gamma(\alpha, \lambda)}$ in the magnitude domain. Depending on the type of the background clutter (i.e., homogeneous, heterogeneous, or extremely heterogeneous), and the pertinent sensor characteristics (i.e., operating conditions) such as frequency, polarization, and gazing angles, several different distributions are used in the literature to model the backscatter $X$. For each case, the manner in which $X$ (and subsequently $Z$) manifests itself depends[31,94] on whether the SAR image is single-look (i.e., $n = 1$) or multilook (i.e., $n > 1$).

First, for homogeneous regions and a single look SAR image, $X$ is typically modeled as a constant that equals the average power in the homogeneous region (i.e., $C = 2\sigma_G^2$). Accordingly, the power-domain SAR image $Z$ is modeled as exponential distributed:

$$Z \sim \exp\left(\frac{\alpha}{\lambda}\right). \tag{34}$$

Similarly, for a multilook SAR image (i.e., $n > 1$), $\alpha = n$, and $\lambda = n/(2\sigma_G^2)$. However, the power domain SAR image $Z$ becomes gamma distributed:

$$Z \sim \Gamma\left(n, n\frac{\lambda}{\alpha}\right). \tag{35}$$

Second, for heterogeneous regions, the backscatter $X$ is not modeled as a constant. Rather, it is modeled as a gamma distribution $\Gamma(\alpha, \lambda)$, or a square root gamma distribution $\Gamma^{1/2}(\alpha, \lambda)$, for power-domain and amplitude-domain SAR imagery, respectively. This yields the $K$-distribution model, $K(\alpha, \lambda, n)$ for any number of looks $n$.

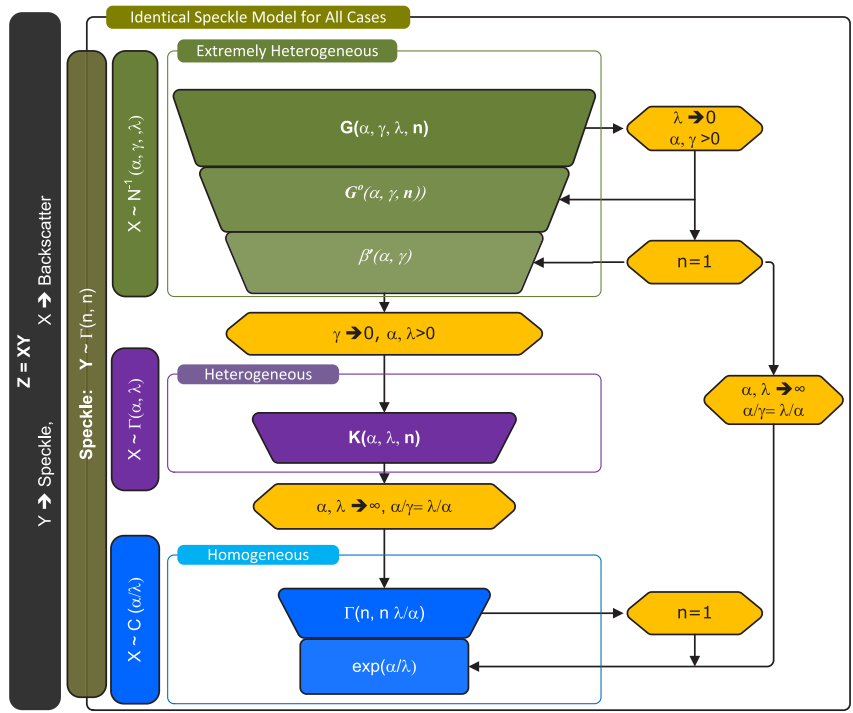

**Fig. 7** The SAR multiplicative (also known as compound) models and interrelations (See Ref. 31.)

Third, for extremely heterogeneous regions, the $G$-distribution[93] is typically used. Unlike the $K$-distribution, the $G$-distribution uses the square root of the generalized inverse Gaussian distribution to model the backscatter $X$ for both homogeneous and heterogeneous backgrounds in the magnitude-domain:

$$X \sim \sqrt{N^{-1}(\alpha, \gamma, \lambda)}. \tag{36}$$

The speckle model is left unchanged as provided earlier. This model is the most generic, and the previous models are special cases of it. Indeed, $\sqrt{N^{-1}(\alpha, \gamma, \lambda)}$ leads to the following three special cases. First, the square root of the gamma distribution leads to the $K$-distribution. Second, the reciprocal of the square root of the gamma distribution leads to the $G^o$-distribution. Third, a constant leads to a scaled speckle (i.e., the homogeneous case), as explained in Eqs. (34) and (35), for single-look and multilook SAR imagery, respectively.

With the same number of free parameters (i.e., two parameters) as the $K$-distribution, the $G^o$-distribution can model extremely heterogeneous regions that the $K$-distribution cannot model. Finally, the $G^o$-distribution reduces to the beta-prime distribution, $\beta'(\alpha, \gamma)$, for single-look (i.e., $n = 1$) SAR imagery.[31] The various multiplicative SAR models and the interrelation between them are summarized in Fig. 7.

## 5.3 CFAR Detection: Two Additional Variant Perspectives

Thus far, we exclusively dealt with the CFAR detection problem from the traditional perspective of the radar community. In this section, we look at CFAR from two additional variant perspectives: a signal processing perspective and a pattern recognition perspective. This section presents a novel discussion and results from these two perspectives. By the end of this section, it will be clear that interpreting CFAR from these additional perspectives opens the door for advancing and improving the CFAR detection process on both the computational complexity (i.e., implementation) level as well as the performance (i.e., false alarm rejection and target detection) level. It should be highlighted that the discussion provided here can be generalized to non-CFAR methods that implement a similar sliding window strategy for neighborhood processing.

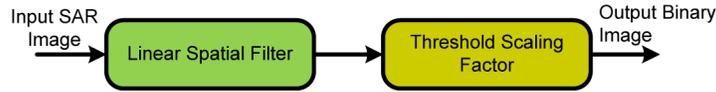

**Fig. 8** CFAR as a linear spatial filter.

### 5.3.1 Signal processing perspective

From a pure signal processing perspective, the sliding window CFAR approach is simply a linear spatial filter for the pixel neighborhoods in the input SAR image. This is followed by an adaptive thresholding operation (i.e., threshold scaling factor) to create a binary image. This process is depicted in Fig. 8.

Recall that, for the log-detector one-parameter CFAR (the subscript log introduced earlier is dropped for notational simplicity), the detection decision is based on

$$\bar{X}_{\text{PUT}} - \hat{\mu}_B \underset{\omega_T}{\overset{\omega_B}{\lessgtr}} \alpha. \qquad (37)$$

Similarly, for the log-detector two-parameter CFAR, the detection decision is given by

$$\frac{\bar{X}_{\text{PUT}} - \hat{\mu}_B}{\hat{\sigma}_B} \underset{\omega_T}{\overset{\omega_B}{\lessgtr}} \alpha. \qquad (38)$$

In these inequalities, $\bar{X}_{\text{PUT}}$, $\hat{\mu}_B$, and $\hat{\sigma}_B$ are estimated based on a linear spatial filtering operation as depicted in Fig. 8. Further, the filtered image is compared with a corresponding matrix pertaining to the threshold scaling factor (i.e., $\alpha$) to produce a binary image of detections. The threshold scaling factor is approximated for each pixel neighborhood (i.e., boundary ring in the CFAR stencil) in the SAR image based on the statistical model chosen for the background clutter and the desired PFA.

Accordingly, the CFAR stencil is a band-pass (BP) finite impulse response (FIR) filter. The BP filter can be designed using two low-pass filters: one centered on the PUT(s), including the guard region, and the other centered on the region correspond to the whole CFAR stencil. This kind of filter is typically referred to as convolution kernel, or simply kernel. We will use the term kernel to refer to the CFAR stencil in the spatial domain.

Prior to elaborating on this, we briefly demonstrate the process of linear spatial filtering and its relation to one-parameter CFAR. We then extend the discussion to two-parameter CFAR.

**Linear spatial filtering, one PUT.** In the general case, an input SAR image $\boldsymbol{I}$ is a pixel array of size $M \times N$. The CFAR kernel $f_B$ has a size $M_B \times N_B$. $M_B$ and $N_B$ are typically chosen to be odd values to avoid ambiguity in defining the PUT. Convolving the SAR image $\boldsymbol{I}$ with the kernel $f_B$ yields

$$\boldsymbol{I}_{\text{Conv},B}(m,n) = f_B * \boldsymbol{I}(m,n) = \sum_{m_B=1}^{M_B} \sum_{n_B=1}^{N_B} f_B(m_B, n_B) \boldsymbol{I}(m - m_B, n - n_B), \qquad (39)$$

where $*$ denotes 2-D convolution, $m \in \{1, 2, \ldots, M_B\}$, and $n \in \{1, 2, \ldots, N_B\}$.

Note that $\boldsymbol{I}_{\text{Conv},B}(m,n)$ is the filtered pixel corresponding to coordinates $(m, n)$. Also note that, because the image is filtered in the spatial domain, the kernel is not multiplied, but rather convolved with the neighborhood pixels. The kernel entries should be chosen to keep the desired boundary ring pixels in the CFAR stencil and eliminate other pixels corresponding to the guard ring and the PUT.

Fig. 9 depicts an illustrative example for a convolution kernel $f_B$. In this example, having only one PUT, one layer guard ring, and one layer boundary ring is desired. Thus, the boundary ring kernel $f_B$ is given by

**Fig. 9** Convolution kernel corresponds to a desired CFAR stencil. Pixels highlighted in green and assigned values of one correspond to the boundary ring. Pixels highlighted in red and black and assigned values of zero correspond to the guard ring and PUT, respectively.

**Fig. 10** Illustration of the convolution process. The black pixels correspond to the original SAR image. The CFAR stencil is systematically slid on the image. (a) $f_B$ (b) $f_{\text{Whole}}$ (c) $f_{\text{PUT+G}}$.

$$f_B = \frac{1}{16} \begin{bmatrix} 1 & 1 & 1 & 1 & 1 \\ 1 & 0 & 0 & 0 & 1 \\ 1 & 0 & 0 & 0 & 1 \\ 1 & 0 & 0 & 0 & 1 \\ 1 & 1 & 1 & 1 & 1 \end{bmatrix}. \tag{40}$$

Following Eq. (39), this filter is convolved with the SAR image $\boldsymbol{I}$ as depicted in Fig. 10.

To illustrate the filtering (i.e., convolution) process, assume that we wish to perform filtering at pixel $x_{71}$ in the original SAR image with the filter as depicted in Fig. 9. The filtering result corresponding to this particular PUT is given by

$$\boldsymbol{I}_{\text{Conv},B}(5, 11) = f_B * \boldsymbol{I}(5, 11)$$

$$= \frac{1}{16}(1 \times x_{39} + 1 \times x_{40} + 1 \times x_{41} + 1 \times x_{42} + 1 \times x_{43} + 1 \times x_{54} + 1 \times x_{58}$$

$$+ 1 \times x_{69} + 1 \times x_{73} + 1 \times x_{84} + 1 \times x_{88} + 1 \times x_{99} + 1 \times x_{100} + 1 \times x_{101}$$

$$+ 1 \times x_{102} + 1 \times x_{103} + 0 + 0 + 0 + 0 + 0 + 0 + 0 + 0 + 0). \tag{41}$$

This process is performed for all the pixels in the original SAR image. Obviously, performing this operation on the whole SAR image yields a filtered (i.e., convolution) image of the same size as the original SAR image. Each pixel in the resultant image represents an ML estimate of the arithmetic average of the pixels in the boundary ring corresponding to the original SAR image.

Note that, depending on the size of the input SAR image in relation to the convolution kernel, the input image may need to be zero-padded prior to performing the convolution.

*Linear spatial filtering, multiple PUTs.* Now consider the more general case where one desires to have more than one PUT in the center of the CFAR stencil. In this case, another kernel $f_T$ with size $M_T \times N_T$ is required. The kernel has a size corresponding to the size of the CFAR stencil. Values of one corresponding to the size of PUTs are inserted in the center of the kernel, and zeros are placed in the guard ring and the boundary ring. Similar to the aforementioned description, the kernel is convolved with the original SAR image as

$$\boldsymbol{I}_{\text{Conv},T}(m,n) = f_T * \boldsymbol{I}(m,n) = \sum_{m_T=1}^{M_T} \sum_{n_T=1}^{N_T} f_T(m_T, n_T) \boldsymbol{I}(m - m_T, n - n_T), \qquad (42)$$

where $*$ denotes 2-D convolution, $m \in \{1, 2, \ldots, M_T\}$, and $n \in \{1, 2, \ldots, N_T\}$, as explained earlier.

*CFAR, one-parameter and two-parameter.* Based on the above-mentioned illustration, for a one-parameter CA-CFAR, the CA-CFAR decision is based on

$$f_T * \boldsymbol{I}(m,n) - f_B * \boldsymbol{I}(m,n) \underset{\omega_T}{\overset{\omega_B}{\lessgtr}} \alpha(m,n), \qquad (43)$$

and for two-parameter CFAR, the decision is based on

$$\frac{f_T * \boldsymbol{I}(m,n) - f_B * \boldsymbol{I}(m,n)}{\hat{\sigma}_B(m,n)} \underset{\omega_T}{\overset{\omega_B}{\lessgtr}} \alpha(m,n), \qquad (44)$$

where $\hat{\sigma}_B(m,n) = [f_B * \boldsymbol{I}(m,n)^2 - (f_B * (\boldsymbol{I}(m,n))^2]^{0.5}$ is the ML estimate of the background variance, and $*$ denotes 2-D convolution.

The adaptive threshold scaling factor $\alpha(m,n)$ for pixel(s) $(m,n)$ under test is calculated based on the probability distribution used to model the background clutter (see Sec. 3.1). This yields a matrix with the same size as the input SAR image (i.e., $M \times N$). Obviously, the kernels $f_T$ and $f_B$ correspond to low-pass filters for the target and boundary ring, respectively, and $f_T * \boldsymbol{I}(m,n) - f_B * \boldsymbol{I}(m,n)$ corresponds effectively to a BP filtering.

*Understanding CFAR as a BP filter.* The typical relationship between spatial domain filtering and frequency domain filtering is given as

$$f_B * \boldsymbol{I}(m,n) \leftrightarrow F_B \mathbb{I}(u,v), \qquad (45)$$

where $*$ denotes 2-D convolution. In this equation, $F_B(u,v)$ and $\mathbb{I}(u,v)$ are the 2-D Fourier transforms of the spatial domain CFAR kernel $f_B$ and the SAR image $I(m,n)$, respectively.

To elaborate further on this equation, let us consider the illustrative CFAR stencil (i.e., kernel) given in Fig. 9 earlier and represented in Fig. 11 hereinafter. Obviously, using the CFAR kernel in Fig. 11(a), the CFAR operation on a SAR image involves some neighborhood operations performed on the boundary ring while discarding the guard ring as well as the PUT (the discussion can be easily generalized to PUTs). The boundary ring is shaded in green in Fig. 11(a). The CFAR kernel can be decomposed into two kernels: one covers the whole CFAR stencil, as shown in Fig. 11(b), while the other covers both the boundary ring and the PUT, as shown in Fig. 11(c).

The spatial kernels in Fig. 11 can be converted to the frequency domain by simply applying the Fourier transform. The magnitude spectra of the 3-D frequency-domain kernels are given in Fig. 12, and their 2-D projections are depicted in Fig. 13.

After the SAR image is BP filtered, the second block in the CFAR processing involves thresholding the filtered image to produce a binary image of detections. This is performed by simply comparing the filtered SAR image from the first stage with a corresponding image (2-D matrix) of thresholds pertaining to each PUT in the image. The matrix of thresholds is typically estimated for a desired PFA based on the desired statistical model for the background clutter.

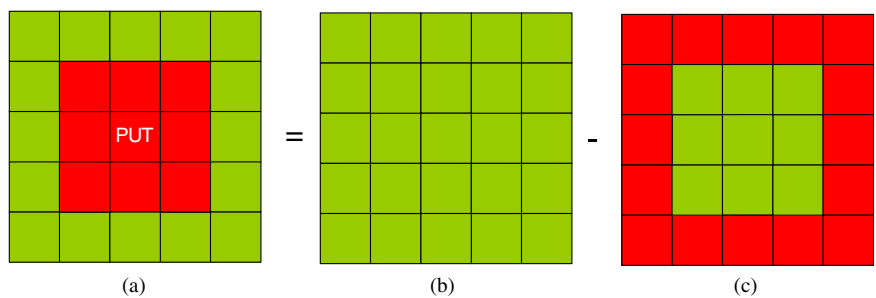

**Fig. 11** CFAR neighborhood kernel ($f_B$) decomposed into two kernels. One kernel covers the whole CFAR stencil ($f_{\text{Whole}}$), and the other covers the guard ring and the PUT ($f_{\text{PUT+G}}$). Pixels shaded in green have a value of one; those shaded in red have a value of zero. (a) $F_B$ (b) $F_{\text{Whole}}$ (c) $F_{\text{PUT+G}}$.

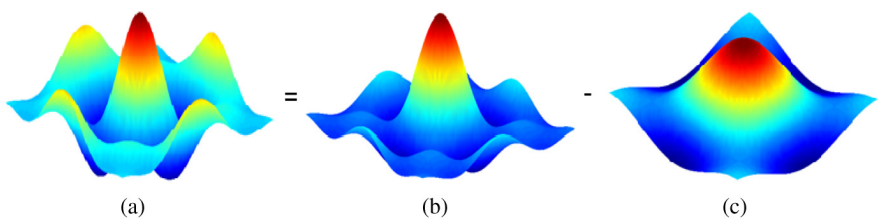

**Fig. 12** 3-D magnitude spectrum for spatial kernels in Fig. 12. (a) $F_B(u, v)$ (b) $F_{\text{Whole}}(u, v)$ (c) $F_{\text{PUT+G}}(u, v)$.

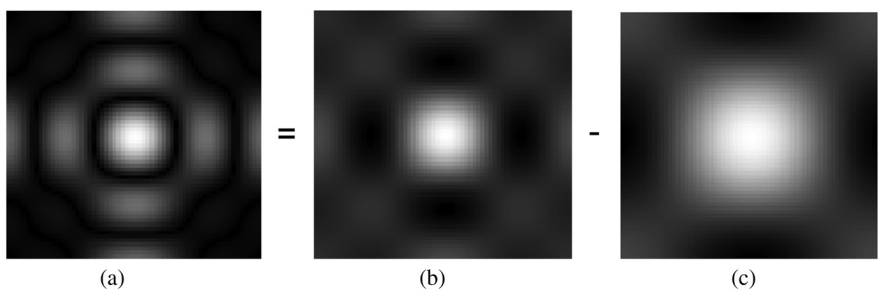

**Fig. 13** Corresponding 2-D magnitude spectrum for Fig. 12.

Our discussion here shows that it is possible to implement the first stage in the CFAR processing either in the spatial domain or in the frequency domain. This choice will depend on the size of the CFAR stencil, and it offers a means for efficient implementation. Typically, for smaller CFAR stencils, it is more efficient to use the spatial implementation, while for larger CFAR stencils, it is more efficient to migrate to the frequency domain implementation.

Another interesting observation is that our understanding of CFAR from the signal processing perspective does not limit us to a rectangle-shaped stencil. Indeed, it is feasible to implement any desired shape for the CFAR stencil. For example, it is feasible to implement an adaptive CFAR stencil that adapts its shape based on some prior information, such as clutter map or image segmentation or differences in 2-D resolution (i.e., range versus azimuth).

In conclusion, this section presented an interesting and novel interpretation for CFAR. This interpretation not only enables an efficient and compact implementation of the CFAR processing, but also enables us to approach the problem from a unique perspective, thus allowing improvements to the detection process potentially to be realized.

### 5.3.2 *Pattern recognition perspective*

Statistical pattern recognition provides a unique interpretation for the CFAR detection process. In this section, the two formulas analyzed in the earlier section for one-parameter and two-parameter (log-detector) CFAR, respectively, are revisited. Then, the well-known quadratic discriminant classifier is presented and rearranged in a way that resembles CFAR. Further, the CFAR formulas are contrasted with the quadratic discriminant classifier. Finally, some insightful conclusions are drawn.

*Log-detector CFAR formulas.* We briefly revisit the log-detector CA-CFAR formulas discussed in Sec. 5.3.1. First, consider the one-parameter CFAR detector. As the (log-detector) CFAR stencil runs over the pixels of the SAR image, the CFAR decision at each PUT(s) for one-parameter CFAR is based on (the subscript log introduced earlier is dropped for notational simplicity):

$$\bar{X}_{\mathrm{PUT}} - \hat{\mu}_B \underset{\omega_T}{\overset{\omega_B}{\gtrless}} \alpha, \tag{46}$$

where $\bar{X}_{\mathrm{PUT}}$ and $\hat{\mu}_B$ are the ML estimates for the arithmetic means of the PUT(s) and the pixels in the boundary ring, respectively, and $\alpha$ is a scaling factor estimated based on the parametric model adopted for the background clutter pertaining to the desired PFA.

Similarly, the log-detector CFAR decision for two-parameter CFAR is given by

$$\frac{\bar{X}_{\mathrm{PUT}} - \hat{\mu}_B}{\hat{\sigma}_B} \underset{\omega_T}{\overset{\omega_B}{\gtrless}} \alpha, \tag{47}$$

where $\hat{\sigma}_B$ is the ML estimate for the standard deviation of the pixels in the boundary ring.

Equations (46) and (47) will be contrasted with the quadratic discriminant classifier discussed below.

*Quadratic discriminant classifier.* In statistical pattern recognition, it is well known that the Bayesian classifier for normal classes can be expressed as a quadratic discriminant function (QDF). Thus, assuming that the detection problem is a binary (i.e., two-class; dichotomizer), the classification entails two discriminant functions:

$$g_{\omega_T}(\boldsymbol{x}) = -\frac{1}{2}(\boldsymbol{x} - \boldsymbol{\mu}_T)^T \Sigma_T^{-1}(\boldsymbol{x} - \boldsymbol{\mu}_T) - \frac{1}{2}\log|\Sigma_T^{-1}| + \log P(\omega_T), \tag{48}$$

$$g_{\omega_B}(\boldsymbol{x}) = -\frac{1}{2}(\boldsymbol{x} - \boldsymbol{\mu}_B)^T \Sigma_B^{-1}(\boldsymbol{x} - \boldsymbol{\mu}_B) - \frac{1}{2}\log|\Sigma_B^{-1}| + \log P(\omega_B), \tag{49}$$

where $g_{\omega_T}(x)$ and $g_{\omega_B}(x)$ are the QDF for the target-class (i.e., $\omega_T$) and the background-class (i.e., $\omega_B$), respectively; $\mu_T$ and $\mu_B$ are the ML mean estimates pertaining to the training features for the target-class and the background-class, respectively; $\Sigma_T^{-1}$ and $\Sigma_B^{-1}$ are the inverse covariance matrices for the target-class and the background-class, respectively; and $P(\omega_{\omega_T})$ and $P(\omega_B)$ are the priors for the target-class and the background-class, respectively.

In the case of CFAR, our feature-space is only one-dimensional. Subsequently, the training mean vector for each class is a 1-D scalar. Moreover, the training covariance matrix for each class reduces to the variance. Further, the test vector is also 1-D, and it is simply the ML estimate of the mean PUTs, $\bar{X}_{\mathrm{PUT}}$. Thus, assuming that CFAR is a binary classification problem, Eqs. (48) and (49) reduce to

$$g_{\omega_T}(\bar{X}_{\mathrm{PUT}}) = -\frac{1}{2}\frac{(\bar{X}_{\mathrm{PUT}} - \mu_T)^2}{\sigma_T^2} - \frac{1}{2}\log|\sigma_T^{-2}| + \log P(\omega_T), \tag{50}$$

$$g_{\omega_B}(\bar{X}_{\mathrm{PUT}}) = -\frac{1}{2}\frac{(\bar{X}_{\mathrm{PUT}} - \mu_B)^2}{\sigma_B^2} - \frac{1}{2}\log|\sigma_B^{-2}| + \log P(\omega_B). \tag{51}$$

Accordingly, the detection decision in this case follows

$$g_{\omega_T}(\bar{X}_{\text{PUT}}) \lessgtr^{\omega_B}_{\omega_T} g_{\omega_B}(\bar{X}_{\text{PUT}}). \tag{52}$$

Subsequently, for the general case, the decision threshold (i.e., decision boundary) between the two classes is given by

$$g_{\omega_T}(\bar{X}_{\text{PUT}}) - g_{\omega_B}(\bar{X}_{\text{PUT}}) = 0. \tag{53}$$

Obviously, this detection decision is a variant from typical CFAR. Indeed, one may reformulate the classification problem from CFAR's perspective (i.e., for the commonly used AD). In this case, one considers only the discriminant function pertinent to the background-class as

$$g_{\omega_B}(\bar{X}_{\text{PUT}}) = -\frac{1}{2}\frac{(\bar{X}_{\text{PUT}} - \mu_B)^2}{\sigma_B^2} - \frac{1}{2}\log|\sigma_B^{-2}| + \log P(\omega_B). \tag{54}$$

Accordingly, one can decide a detection decision as

$$-g_{\omega_B}(\bar{X}_{\text{PUT}}) \lessgtr^{\omega_B}_{\omega_T} \alpha_{\text{QDF}}, \tag{55}$$

where $\alpha_{\text{QDF}}$ is a scaling factor pertaining to the detection threshold. Thus, the detection decision is based on

$$\frac{1}{2}\frac{(\bar{X}_{\text{PUT}} - \mu_B)^2}{\sigma_B^2} + \frac{1}{2}\log|\sigma_B^{-2}| - \log P(\omega_B) \lessgtr^{\omega_B}_{\omega_T} \alpha_{\text{QDF}}. \tag{56}$$

Obviously, this is a suboptimal one-class classification problem.[23] It is stated in Principe et al.[20] that CFAR as an anomaly detector is a two-class classification problem. Indeed, this statement is incorrect. CFAR formulates a discriminant function for a (sub-optimal) one-class classification as illustrated in Eqs. (55) and (56).

*Contrasting CA-CFAR with QDF.*   Let us first compare the one-parameter CFAR, as shown in Eq. (46), with the QDF, as shown in Eq. (56). Squaring both sides of Eq. (46) yields

$$(\bar{X}_{\text{PUT}} - \hat{\mu}_B)^2 \lessgtr^{\omega_B}_{\omega_T} \alpha^2. \tag{57}$$

This equation equals to Eq. (56) when $\sigma_{\omega_B}^2 = 1$, and $2[\alpha_{\text{QDF}} + \log P(\omega_B)] = \alpha^2$. Subsequently, the one-parameter CFAR is simply a (squared) Euclidean distance classifier. More precisely, under this case, the QDF reduces to a linear minimum distance classifier based on the Euclidean distance.

Similarly, this discussion can be extended to the two-parameter CFAR. First, squaring both sides of Eq. (47) yields

$$\frac{(\bar{X}_{\text{PUT}} - \hat{\mu}_B)^2}{\hat{\sigma}_B} \lessgtr^{\omega_B}_{\omega_T} \alpha^2. \tag{58}$$

Comparing this inequality with Eq. (56) one notes the following. First, the scaling factor in the two-parameter CFAR equation corresponds to

$$\alpha^2 = 2\alpha_{\text{QDF}} + 2\log P(\omega_B). \tag{59}$$

Second, the term $\log|\sigma_{\omega_B}^{-2}|$ is missing from the equation of two-parameter CFAR, Eq. (58). Subsequently, the conclusion is reached that the two-parameter CFAR is a quadratic discriminant classifier applied to the logarithm of the power (i.e., magnitude-squared) SAR image. However, $\log|\sigma_{\omega_B}^{-2}|$ is missing, nonetheless.

Subsequently, from the perspective of optimal classification (i.e., Bayes classifier), under the assumption that all conditions for optimality are fulfilled, two-parameter CFAR (as an anomaly detector) can never achieve optimal performance, due to the missing term in the above equation (practically speaking). Further, the online training methodology based on the pixel neighborhoods in the log-domain SAR image is not efficient, because the neighborhoods are chosen *ad hoc*. Note that our conclusions on two-parameter CFAR resemble those published in Principe, Kim, and Fisher.[67] However, unlike our conclusions, which are based on CFAR as a log-detector, the conclusions in Principe, Kim, and Fisher are based on an envelope (i.e., magnitude) CFAR detector. Their assumption of the magnitude detection is inaccurate, because the formula for two-parameter CFAR used here and in their work was designed for log detection.

In conclusion, this section demonstrates the relationship between CFAR and the quadratic discriminant classifier. Indeed, regardless of the various models used in the literature to estimate the scaling factor of the detection threshold (Weibull distribution, $K$-distribution, etc.) our discussion makes it clear that both one-parameter CFAR and two-parameter CFAR are based on the inherent assumption that the class-likelihoods are normal. Under the assumption that CFAR is an anomaly detector, the CFAR classifier was shown to be a suboptimal one-class classifier. One-parameter CFAR was shown to be a special case of the quadratic discriminant classifier, a linear minimum distance classifier. More precisely, it is a Euclidean distance classifier. Two-parameter CFAR was shown to be a quadratic discriminant classifier but with a missing term that adds to the performance degradation. These interesting insights enable one to understand the inherent limitations of CFAR, and they pave the way for future improvements.

## 6 Conclusions

An end-to-end ATR system for SAR imagery (SAR-ATR) is typically multistaged to counter the prohibitive amounts of processing pertinent to the input SAR imagery. The front-end stage in any such SAR-ATR system is the detector (also known as prescreener). This paper has presented a state-of-the-art survey of the various methods for target detection in SAR imagery. First, the major methods reported in the open literature for implementing the detector are broadly taxonomized into single-feature-based, multifeature-based, and expert-system-oriented. Then, various implementation strategies under each taxon are further classified and overviewed. Special attention is paid to the Bayesian approach being the optimal approach. Additionally, emphasis is placed on the CFAR approach being the most popular. Further, the obvious advantages of the expert-system-oriented taxon are noted. Selections of representative examples from the literature under each method are presented. A table of comparison between selected methods under each taxon is provided. Finally, we elaborate on a novel discussion on important issues pertaining to target detection in SAR imagery.

It is shown that the beta-prime ($\beta'$) distribution and $G^o$-distribution allow for better means (compared to the $K$-distribution) for modeling the background clutter for single-look and multi-look SAR imagery, respectively. Further, CFAR is analyzed from two interesting perspectives: signal processing and statistical pattern recognition. From the signal processing perspective, CFAR is shown to be a FIR BP filter that can be readily implemented using a two suitable LP filters. This implementation can be realized either in the spatial domain or in the frequency domain. From the perspective of statistical pattern recognition, the anomaly detector CFAR is shown to be a suboptimal one-class classifier. One-parameter CFAR is found to be a Euclidean distance classifier, whereas two-parameter CFAR is shown to be a quadratic discriminant with a missing term that adds to the CFAR loss. These interpretations of CFAR allow for potential future improvements on the design level, as well as the implementation level.

It should be noted that this paper did not aim at providing an exhaustive survey for all the available methods in the open literature to implement a detector. This is infeasible, due to the volume of works published on the topic. Rather, this paper has taxonomized the major approaches and provided a careful selection of popular methods along with representative examples pertaining to SAR imagery. Additional methods unreported in this paper fall under one of the taxa provided. Future work plans will consider investigating a proper method for context acquisition and utilization in SAR imagery for use in an expert-system-oriented detection scheme. This will be appropriately tailored to Radarsat-2 imagery acquired in the Spotlight mode. This is the underlying motivation that drove the development of this survey. Further, a comprehensive state-of-the-art survey pertaining to the intermediate and back-end stages in the SAR-ATR processing chain is under preparation.

## Acknowledgments

This work is supported in part by the Research and Development Corporation (RDC) of Newfoundland and Labrador, the Atlantic Innovation Fund (AIF), and the Natural Sciences and Engineering Research Council of Canada (NSERC). The authors would like to thank Dr. Paris Vachon from Defence R&D Canada (DRDC Ottawa), who provided advice and oversight to the project team for the last three years. The first author would like to acknowledge Dr. Cecile Badenhorst from the Faculty of Education at Memorial University of Newfoundland for her inspiring approach towards research as delivered in her workshop Thinking Creatively About Research conducted in the winter semester of 2012, which aided him in the conception of this paper. The authors also thank the editor and anonymous referees for their insightful comments, which improved this paper.

## References

1. D. E. Kreithen, S. D. Halversen, and G. J. Owirka, "Discriminating targets from clutter," *Lincoln Lab. J.* **6**(1), 25–52 (1993).
2. L. M. Novak et al., "Effects of polarization and resolution on SAR ATR," *IEEE Trans. Aerosp. Electron. Syst.* **33**(1), 102–116 (1997), http://dx.doi.org/10.1109/7.570713.
3. L. M. Novak et al., "The automatic-target recognition system in SAIP," *Lincoln Lab. J.* **10**(2), 187–202 (1997).
4. L. M. Novak, G. J. Owirka, and W. S. Brower, "An efficient multi-target SAR ATR algorithm," in *Proc. 32nd Asilomar Conf. on, Signals, Syst. and Comput.*, Vol. 1, pp. 3–13, IEEE, Pacific Grove, California (1998).
5. T. Cooke, "Detection and classification of objects in synthetic aperture radar imagery," Intelligence, Surveillance and Reconnaissance Division Information Sciences Laboratory, Defense Science and Technology Organization, Department of Defense, Australian Government, DSTO–RR–0305 (2006), http://handle.dtic.mil/100.2/ADA449712.
6. H. Chiang and R. L. Moses, "ATR performance prediction using attributed scattering features," *Proc. SPIE* **3721**, 785–796 (1999), http://dx.doi.org/10.1117/12.357693.
7. N. M. Sandirasegaram, "Spot SAR ATR using wavelet features and neural network classifier," *DRDC Ottawa TM 2005-154. R&D pour la défense Canada—Ottawa* (2005), http://pubs.drdc.gc.ca/PDFS/unc89/p524497.pdf.
8. J. Schroeder, "Automatic target detection and recognition using synthetic aperture radar imagery," presented at *Workshop on the Applications of Radio Science, WARS02* (2002).
9. L. M. Kaplan, R. Murenzi, and K. R. Namuduri, "Extended fractal feature for first-stage SAR target detection," *Proc. SPIE* **3721**, 35–46 (1999), http://dx.doi.org/10.1117/12.357684.
10. MacDonald, Dettwiler and Associates Ltd., Dettwiler and Associates Ltd., "Radarsat-2 product description," *MDA Corporation*, Issue 1/9, RN-SP-52-1238 (2011), http://gs.mdacorporation.com/includes/documents/RN-SP-52-1238_RS2_Product_Description_Iss1-9.pdf.
11. J. Lee and E. Pottier, *Polarimetric Radar Imaging From Basics to Applications*, CRC Press, Taylor and Francis Group, Boca Raton, Florida (2009).
12. R. Hansch and O. Hellwich, Chapter 5, in *Radar Remote Sensing of Urban Areas*, Springer, London, New York (2010).
13. R. Touzi et al., "A review of polarimetry in the context of synthetic aperture radar: concepts and information extraction," *Can. J. Remote Sens.* **30**(3), 380–407 (2004), http://dx.doi.org/10.5589/m04-013.
14. L. M. Novak et al., "Optimal polarimetric processing for enhanced target detection," in *Proc. IEEE Telesystems Conf.*, pp. 69–75, IEEE, Atlanta, Georgia (1991).

15. S. R. Cloude and E. Pottier, "A review of target decomposition theorems in radar polarimetry," *IEEE Trans. Geosci. Remote Sens.* **34**(2), 498–518 (1996), http://dx.doi.org/10.1109/TGRS.2008.2006504.

16. L. Zhang et al., "Comparison of methods for target detection and applications using polarimetric SAR image," *PIERS Online* **4**(1), 140–145 (2008), http://dx.doi.org/10.2529/PIERS070907024917.

17. L. M. Novak et al., "Optimal processing of polarimetric synthetic-aperture radar imagery," *Lincoln Lab. J.* **3**(2), 273–290 (1990).

18. R. O. Duda, P. E. Hart, and D. G. Stork, *Pattern Classification*, 2nd ed., Wiley-Interscience, New York (2000).

19. L. M. Novak, G. Owirka, and C. M. Netishen, "Performance of a high-resolution polarimetric SAR automatic target recognition system," *Lincoln Lab. J.* **6**(1), 11–24 (1993).

20. J. C. Principe et al., "Target prescreening based on a quadratic gamma discriminator," *IEEE Trans. Aerosp. Electron. Syst.* **34**(3), 706–715 (1998), http://dx.doi.org/10.1109/7.705880.

21. Array Systems Computing Inc., "Object detection," *NEST*, (3 May 2012), http://www.array.ca/nest-web/help/operators/ObjectDetection.html.

22. D. J. Crisp, "The state-of-the-art in ship deteciton in synthetic aperture radar imagery," Intellegence, Surveillance and Reconnaissance Division Information Sciences Laboratory, Defense Science and Technology Organization, Department of Defense, Australian Government, DSTO-RR-0272 (2004), http://dspace.dsto.defence.gov.au/dspace/bitstream/1947/3354/1/DSTO-RR-0272%20PR.pdf.

23. D. J. Tax, "One-class classification: Concept-learning in the absence of counter-examples," Doctoral Dissertation, Technische Universiteit Delft (2001).

24. D. Blacknell and R. J. A. Tough, "Clutter discrimination in polarimetric SAR imagery," *Proc. SPIE* **2584**, 188–199 (1995), http://dx.doi.org/10.1117/12.227127.

25. D. Blacknell and R. J. A. Tough, "Clutter discrimination in polarimetric and interferometric synthetic aperture radar imagery," *J. Phys. D* **30**(4), 551–556 (1997), http://dx.doi.org/10.1088/0022-3727/30/4/009.

26. M. Di Bisceglie and C. Galdi, "CFAR detection of extended objects in high-resolution SAR images," *IEEE Trans. Geosci. Remote Sens.* **43**(4), 833–843 (2005), http://dx.doi.org/10.1109/TGRS.2004.843190.

27. M. Di Bisceglie and C. Galdi, "CFAR detection of extended objects in high resolution SAR images," *IEEE Geosci. Remote Sens. Symp.* **6**(1), 2674–2676 (2001), http://dx.doi.org/10.1109/IGARSS.2001.978126.

28. S. Kuttikkad and R. Chellappa, "Non-Gaussian CFAR techniques for target detection in high resolution SAR images," in *Proc. IEEE Int. Conf. Image Process.*, Vol. 1, pp. 910–914, IEEE, Austin, Texas (1994).

29. M. Liao et al., "Using SAR images to detect ships from sea clutter," *IEEE Geosci. Remote Sens. Lett.* **5**(2), 194–198 (2008), http://dx.doi.org/10.1109/LGRS.2008.915593.

30. J. Xu et al., "Small target detection in SAR image using the alpha-stable distribution model," in *Proc. IEEE Int. Conf. on Image Analysis and Signal Process.*, pp. 64–68, IEEE, Zhejiang, China (2010).

31. J. S. Salazar, "Detection schemes for synthetic aperture radar imagery based on a beta prime statistical model," Doctoral Dissertation, University of New Mexico (1999).

32. L. M. Novak and S. R. Hesse, "On the performance of order-statistics CFAR detectors," in *Proc. 25th IEEE Asilomar Conf. on Signals, Syst. and Comput.*, Vol. 1, pp. 835–840, IEEE, Pacific Grove, California (1991).

33. J. A. Ritcey and H. Du, "Order statistic CFAR detectors for speckled area targets in SAR," in *Proc. 25th IEEE Asilomar Conf. on Signals, Syst. and Comput.*, Vol. 2, pp. 1082–1086, IEEE, Pacific Grove, California (1991).

34. J. S. Salowe, "Very fast SAR detection," *Proc. SPIE* **2755**, 58–69 (1996), http://dx.doi.org/10.1117/12.243204.

35. R. Gan and J. Wang, "Distribution-based CFAR detectors in SAR images," *J. Syst. Eng. Electron.* **17**(4), 717–721 (2006), http://dx.doi.org/10.1016/S1004-4132(07)60004-8.

36. G. Rong-Bing and W. Jian-Guo, "Distribution-based CFAR detection in SAR images," *J. Syst. Eng. Electron.* **17**(4), 717–721 (2006), http://dx.doi.org/10.1016/S1004-4132 (07)60004-8.

37. H. M. Finn and R. S. Johnson, "Adaptive detection mode with threshold control as a function of spatial sampled clutter level estimates," *RCA Rev.* **29**(1), 414–464 (1968).

38. R. Viswanathan, "Order statistics application to CFAR radar target detection," Chapter 23, in *Handbook of Statistics*, N. Balakrishnan and C. R. Rao, Eds., Vol. 17, pp. 643–671, Elsevier, Amsterdam, The Netherlands (1998).

39. H. You et al., "A new CFAR detector based on ordered statistics and cell averaging," in *Proc. CIE Int. Conf. of Radar*, pp. 106–108, IEEE, Beijing. China (1996).

40. M. A. Richards, *Fundamentals of Radar Signal Processing*, McGraw-Hill Professional, New York (2005).

41. G. V. Trunk, "Range resolution of targets using automatic detectors," *IEEE Trans. Aerosp. Electron. Syst.* **AES-14**(5), 750–755 (1978), http://dx.doi.org/10.1109/TAES.1978 .308625.

42. H. Rohling, "Radar CFAR thresholding in clutter and multiple target situations," *IEEE Trans. Aerosp. Electron. Syst.* **AES-19**(4), 608–621 (1983), http://dx.doi.org/10.1109/ TAES.1983.309350.

43. M. F. Rimbert and R. M. Bell, "Multiresolution order-statistic CFAR techniques for radar target detection," *Proc. SPIE* **5674**, 282–292 (2005), http://dx.doi.org/10.1117/12.600577.

44. M. R. Bell and M. F. Rimbert, "Constant false alarm rate detection techniques based on empirical distribution function statistics," Doctoral Dissertation, Purdue University (2005).

45. P. P. Gandhi and S. A. Kassam, "Analysis of CFAR processors in non-homogeneous background," *IEEE Trans. Aerosp. Electron. Syst.* **24**(4), 427–445 (1988), http://dx.doi.org/ 10.1109/7.7185.

46. L. M. Novak and S. R. Hesse, "On the performance of order-statistics CFAR detectors," in *Proc. 25th IEEE Asilomar Conf. on Signals, Syst. and Comput.*, Vol. 2, pp. 835–840, IEEE, Pacific Grove, California (1991).

47. A. Tom and R. Viswanathan, "Switched order statistics CFAR test for target detection," in *Proc. IEEE Radar Conf.*, pp. 1–5, IEEE, Rome, Italy (2008).

48. R. Viswanathan and A. Eftekhari, "A selection and estimation test for multiple target detection," *IEEE Trans. Aerosp. Electron. Syst.* **28**(2), 509–519 (1992), http://dx.doi.org/ 10.1109/7.144576.

49. T.-T. Van Cao, "A CFAR thresholding approach based on test cell statistics," in *Proc. IEEE Radar Conf.*, pp. 349–354, IEEE, Rome, Italy (2004).

50. L. Tabet and F. Soltani, "A generalized switching CFAR processor based on test cell statistics," *Signal Image Video Process.* **3**(3), 265–273 (2009), http://dx.doi.org/ 10.1007/s11760-008-0075-2.

51. P. P. Gandhi and S. A. Kassam, "Optimality of the cell averaging CFAR detector," *IEEE Trans. Inform. Theor.* **40**(4), 1226–1228 (1994), http://dx.doi.org/10.1109/18 .335950.

52. D. Schleher, "Harbor surveillance radar detection performance," *IEEE J. Ocean. Eng.* **2**(4), 318–325 (1977), http://dx.doi.org/10.1109/JOE.1977.1145358.

53. G. B. Goldstein, "False-alarm regulation in log-normal and weibull clutter," *IEEE Trans. Aerosp. Electron. Syst.* **AES-9**(1), 84–92 (1973), http://dx.doi.org/10.1109/TAES.1973 .309705.

54. F. Meyer and S. Hinz, "Automatic ship detection in space-borne SAR imagery," *Int. Arch. Photogram. Remote Sens. Spatial Inform. Sci.* **38**(1), Part 1-4-7/W5 (2009).

55. C. Oliver and S. Quegan, *Understanding Synthetic Aperture Radar Images*, SciTech Publishing, Raleigh, North Carolina (2004).

56. D. K. Barton, *Modern Radar System Analysis*, Artech House, Norwood, Massachusetts (1988).

57. I. Antipov and J. Baldwinson, "Estimation of a constant false alarm rate processing loss for a high-resolution maritime radar system," Electronic Warfare and Radar Division Defense Science and Technology Organization, Department of Defense, Australian

Government, DSTO-TR-2158 (2008), http://www.dtic.mil/cgi-bin/GetTRDoc?AD=ADA491871.

58. V. G. Hansen and H. R. Ward, "Detection performance of the cell averaging LOG/CFAR receiver," *IEEE Trans. Aerosp. Electron. Syst.* **AES-8**(5), 648–652 (1972), http://dx.doi.org/10.1109/TAES.1972.309580.

59. L. M. Novak, "Radar target detection and map-matching algorithm studies," *IEEE Trans. Aerosp. Electron. Syst.* **AES-16**(5), 620–625 (1980), http://dx.doi.org/10.1109/TAES.1980.308928.

60. F. E. Nathanson, *Radar Design Principles—Signal Processing and the Environment*, 2nd ed., McGraw-Hill, Mendham, New Jersey (1999).

61. B. C. Armstrong and H. D. Griffiths, "CFAR detection of fluctuating targets in spatially correlated *K*-distributed clutter," *Proc. IEEE Radar Signal Process.* **138**(2), 139–152 (1991).

62. G. Gao et al., "An adaptive and fast CFAR algorithm based on automatic censoring for target detection in high-resolution SAR images," *IEEE Trans. Geosci. Remote Sens.* **47**(6), 1685–1697 (2009), http://dx.doi.org/10.1109/TGRS.2008.2006504.

63. R. J. Sullivan, *Radar Foundations for Imaging and Advanced Concepts*, 1st ed., SciTech Publishing, Raleigh, North Carolina (2004).

64. C. M. Bishop, *Pattern Recognition and Machine Learning*, 1st ed., Springer-Verlag, New York (2006).

65. G. Gao, "A parzen-window-kernel-based CFAR algorithm for ship detection in SAR images," *IEEE Geosci. Remote Sens. Lett.* **8**(3), 557–561 (2011), http://dx.doi.org/10.1109/LGRS.2010.2090492.

66. J. C. Principe et al., "Target prescreening based on 2D gamma kernels," *Proc. SPIE* **2487**, 251–258 (1995), http://dx.doi.org/10.1117/12.210842.

67. J. C. Principe, M. Kim, and M. Fisher III, "Target discrimination in synthetic aperture radar using artificial neural networks," *IEEE Trans. Image Process.* **7**(8), 1136–1149 (1998), http://dx.doi.org/10.1109/83.704307.

68. K. Ouchi et al., "Ship detection based on coherence images derived from cross correlation of multilook SAR images," *IEEE Geosci. Remote Sens. Lett.* **1**(3), 184–187 (2004), http://dx.doi.org/10.1109/LGRS.2004.827462.

69. D. Howard, S. Roberts, and R. Brankin, "Target detection in SAR imagery by genetic programming," *Adv. Eng. Software* **30**(5), 303–311 (1999), http://dx.doi.org/10.1016/S0965-9978(98)00093-3.

70. E. Conte, M. Lops, and G. Ricci, "Radar detection in *K*-distributed clutter," *IEE Proc. Radar Sonar Navigat.* **141**(2), 116–118 (1994), http://dx.doi.org/10.1049/ip-rsn:19949882.

71. L. M. Kaplan, "Improved SAR target detection via extended fractal features," *IEEE Trans. Aerosp. Electron. Syst.* **37**(2), 436–451 (2001), http://dx.doi.org/10.1109/7.937460.

72. Q. H. Pham, T. M. Brosnan, and J. M. Smith, "Multistage algorithm for detection of targets in SAR image data," *Proc. SPIE* **3070**, 66–75 (1997), http://dx.doi.org/10.1117/12.281583.

73. N. S. Subotic et al., "Multiresolution detection of coherent radar targets," *IEEE Trans. Image Process.* **6**(1), 21–35 (1997), http://dx.doi.org/10.1109/83.552094.

74. M. Tello, C. Lopez-Martinez, and J. J. Mallorqui, "A novel algorithm for ship detection in SAR imagery based on the wavelet transform," *IEEE Geosci. Remote Sens. Lett.* **2**(2), 201–205 (2005), http://dx.doi.org/10.1109/LGRS.2005.845033.

75. J.-P. Antoine et al. , *Two-Dimensional Wavelets and Their Relatives*, Cambridge University Press, Cambridge, United Kingdom (2004).

76. T. Tao et al., "Targets detection in SAR image used coherence analysis based on S-transform," Chapter 1, in *Electrical Engineering and Control*, M. Zhu, Ed., Vol. 98, pp. 1–9, Springer, Berlin, Heidelberg  (2011).

77. V. C. Chen and H. Ling, *Time Frequency Transforms for Radar Imaging and Signal Analysis*, Artech House, Norwood, Massachusetts (2002).

78. S. Haykin and T. Bhattacharya, "Wigner-Ville Distribution: An Important Functional Block for Radar Target Detection in Clutter," in *Proc. 25th IEEE Asilomar Conf.*

on *Signals, Syst. and Comput.*, Vol. 1, pp. 68–72, IEEE, Pacific Grove, California (1994).

79. I. McConnell and C. Oliver, "Segmentation-based target detection in SAR," *Proc. SPIE* **3869**, 45–54 (1999), http://dx.doi.org/10.1117/12.373158.

80. M. E. Smith and P. K. Varshney, "VI-CFAR: a novel CFAR algorithm based on data variability," in *Proc. IEEE National Radar Conf.*, pp. 263–268, IEEE, Syracuse, New York (1997).

81. M. E. Smith and P. K. Varshney, "Intelligent CFAR processor based on data variability," *IEEE Trans. Aerosp. Electron. Syst.* **36**(3), 837–847 (2000), http://dx.doi.org/10.1109/7.869503.

82. G. T. Capraro et al., "Artificial intelligence and sensor fusion," in *Proc. IEEE Int. Conf. on Integration of Knowledge Intensive Multi-Agent Systems*, pp. 591–595, IEEE, Utica, New York (2003).

83. C. M. Wicks Jr., W. J. Baldygo, and R. D. Brown, "Expert system constant false alarm rate (CFAR) processor," U.S. Patent No. 5,499,030 (1996).

84. W. Baldygo et al., "Artificial intelligence applications to constant false alarm rate (CFAR) processing," in *Proc. IEEE National Radar Conf.*, pp. 275–280, IEEE, Lynnfield, Massachusetts (1993).

85. Y. Cui, J. Yang, and Y. Yamaguchi, "CFAR ship detection in SAR images based on log-normal mixture models," in *Proc. 3rd IEEE Int. Asia-Pacific Conf. on Synthetic Aperture Radar*, pp. 1–3, IEEE, Seoul, South Korea (2011).

86. M. Guida, M. Longo, and M. Lops, "Biparametric linear estimation for CFAR against Weibull clutter," *IEEE Trans. Aerosp. Electron. Syst.* **28**(1), 138–151 (1992), http://dx.doi.org/10.1109/7.135440.

87. F. T. Ulaby and M. C. Dobson, *Handbook of Radar Scattering Statistics for Terrain*, Artech House, Norwood, Massachusetts (1988).

88. A. L. Maffett and C. C. Wackerman, "The modified beta density function as a model for synthetic aperture radar clutter statistics," *IEEE Trans. Geosci. Remote Sens.* **29**(2), 277–283 (1991), http://dx.doi.org/10.1109/36.73669.

89. J. Jao, "Amplitude distribution of composite terrain radar clutter and the $\kappa$-distribution," *IEEE Trans. Antennas Propag.* **32**(10), 1049–1062 (1984), http://dx.doi.org/10.1109/TAP.1984.1143200.

90. E. Jakeman and P. Pusey, "A model for non-Rayleigh sea echo," *IEEE Trans. Antennas Propag.* **24**(6), 806–814 (1976), http://dx.doi.org/10.1109/TAP.1976.1141451.

91. C. F. Yanasse et al., "On the use of multilook amplitude K distribution for SAR image analysis," in *Proc. IEEE Int. Geosci. and Remote Sens. Symp., Surface and Atmospheric Remote Sens.: Technologies, Data Analysis and Interpretation*, Vol. 4, pp. 2173–2175, IEEE, Pasadena, California (1994).

92. L. M. Novak, "K-distribution and polarimetric terrain radar clutter," *J. Electromagn. Waves Appl.* **3**(8), 747–768 (1989), http://dx.doi.org/10.1163/156939389X00412.

93. A. C. Frery et al., "A model for extremely heterogeneous clutter," *IEEE Trans. Geosci. Remote Sens.* **35**(3), 648–659 (1997), http://dx.doi.org/10.1109/36.581981.

94. A. C. Frery et al., "Models for synthetic aperture radar image analysis," *Resenhas (IME-USP)* **4**(1), 45–77 (1999).

95. Y. Dong, "Models of land clutter vs grazing angle, spatial distribution and temporal distribution—L-band VV polarisation perspective," Electronic Warfare and Radar Division Defense Science and Technology Organization, Department of Defense, Australian Government, DSTO-RR-0273 (2004).

96. L. J. Marier Jr., "Correlated *K*-distributed clutter generation for radar detection and track," *IEEE Trans. Aerosp. Electron. Syst.* **31**(2), 568–580 (1995), http://dx.doi.org/10.1109/7.381906.

97. R. B. D'Agostino and M. A. Stephens, *Goodness-of-Fit Techniques*, 1st ed., Dekker, New York (1986).

98. M. A. Stephens, "EDF statistics for goodness of fit and some comparisons," *J. Am. Stat. Assoc.* **69**(347), 730–737 (1974), http://dx.doi.org/10.1080/01621459.1974.10480196.

99. F. W. Scholz and M. A. Stephens, "K-sample Anderson-Darling tests," *J. Am. Stat. Assoc.* **82** (399), 918–924 (1987).
100. K. P. Brunham and D. R. Anderson, *Model Selection and Multi-Model Inference: A Practical Information-Theoretic Approach*, Springer, New York (2010).
101. J. C. Dainty, *Laser Speckle and Related Phenomena*, Springer, Berlin (1975).

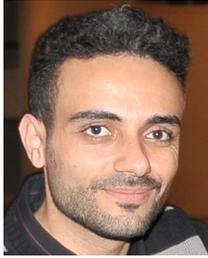

**Khalid El-Darymli** is a PhD candidate in the Faculty of Engineering and Applied Science at Memorial University of Newfoundland, Canada. He is also affiliated with C-CORE. He received his MSc with distinction in computer and information engineering from the International Islamic University of Malaysia and his BSc in electrical engineering with honors from Garyounis University of Libya. He is a recipient of the Ocean Industries Student Research Award from the Research & Development Corporation in Newfoundland, Canada. His current research interests include synthetic aperture radar, target detection, and automatic target recognition in SAR imagery. He is a member of the Association of Professional Engineers and Geoscientists of Alberta.

**Peter McGuire** holds a BASc (engineering) and a PhD in aerospace engineering from the University of Toronto, where he studied the use of artificial neural networks for computer vision and control of dynamic systems. Since joining C-CORE, he has specialized in image processing, earth observation, and data fusion projects. Projects include earth observation using a virtual SAR constellation of satellites, sensor management and data fusion system design using holonic control, along with high-speed automated inspection using computer vision. In addition to project-related activities, he is cross-appointed with Memorial University of Newfoundland and manages a research program focused on oil- and gas-related issues. Topics include detection and mapping of oil under ice using autonomous underwater vehicles, advanced techniques for monitoring targets and infrastructure using satellite and ground-based SAR, coordination of aerial and ground-based robotic systems, and sense and avoid algorithms for unmanned aerial vehicles.

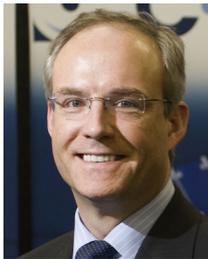

**Desmond Power** MEng, PEng, is vice president of remote sensing at C-CORE. He has more than 21 years of experience in radar and remote sensing. He started his career in terrestrial radar, working as an RF designer on an over-the-horizon radar. He was also heavily involved with signal processing and analysis of radar data. Soon after the launch of RADARSAT in 1995, he moved into projects related to earth observation, with his first project dealing with iceberg detection capabilities of SAR. Since that time, he has managed and been technical advisor to a large series of projects at C-CORE involving earth observation, including marine target detection, vehicle detection along right-of-ways, and interferometry for ground deformation measurement. He still is actively involved in development of terrestrial-based radar systems. He is presently the PI of a multimillion-dollar R&D project on radar-based critical infrastructure monitoring funded by the Atlantic Innovation Fund. He is a member of the IEEE and the Association of Professional Engineers and Geoscientists of Newfoundland and Labrador.

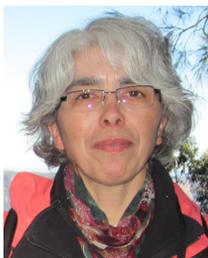

**Cecilia Moloney** received a BSc (honors) degree in mathematics from Memorial University of Newfoundland, Canada, and MASc and PhD degrees in systems design engineering from the University of Waterloo, Canada. Since 1990, she has been a faculty member with Memorial University, where she is now a full professor of electrical and computer engineering. From 2004 to 2009, she held the NSERC/Petro-Canada Chair for Women in Science and Engineering, Atlantic Region. Her research interests include nonlinear signal and image processing methods, signal representations via wavelet and contourlet transforms, radar signal processing, transformative pedagogy for science and engineering, and gender and science studies.